\documentclass[aps,pre,preprint]{revtex4}
\usepackage{appendix}
\usepackage{epsfig}
\usepackage{amsfonts}
\usepackage{amsmath}
\usepackage{psfrag}
\usepackage{amssymb,amsmath,amsthm}
\usepackage{verbatim}
\usepackage{soul}
\usepackage[normalem]{ulem}
\usepackage{color}

\usepackage{amsfonts}
\usepackage{amsmath}
\usepackage{psfrag}
\usepackage{amssymb,amsmath,amsthm}
\usepackage{color}
\newcommand{\beq}{\begin{equation}}
\newcommand{\eeq}{\end{equation}}
\newcommand{\bea}{\begin{eqnarray}}
\newcommand{\eea}{\end{eqnarray}}

\newcommand{\bma}{\left(\begin{matrix}}
\newcommand{\ema}{\end{matrix}\right)}

\newcommand{\om}{\omega}
\begin{document}


\title{Diffusion of elastic waves in a two dimensional continuum with a random distribution of screw dislocations}

\author{Dmitry Churochkin and Fernando Lund}

\affiliation{Departamento de F\'\i sica and CIMAT, Facultad de Ciencias
F\'\i sicas y Matem\'aticas, Universidad de Chile, Santiago, Chile}

\begin{abstract}
We study the diffusion of anti-plane elastic waves in a two dimensional continuum by many, randomly placed, screw dislocations. Building on a previously developed theory for coherent propagation of such waves, the incoherent behavior is characterized by way of a Bethe Salpeter (BS) equation. A Ward-Takahashi identity (WTI) is demonstrated  and the BS equation is solved, as an eigenvalue problem, for long wavelengths and low frequencies. A diffusion equation results and the diffusion coefficient $D$ is calculated. The result has the expected form $D = v^* l /2$, where $l$, the mean free path, is equal to the attenuation length of the coherent waves propagating in the medium and the transport velocity is given by $v^*= c_T^2/v$, where $c_T$ is the wave speed in the absence of obstacles and $v$ is the speed of coherent wave propagation in the presence of dislocations.
\end{abstract}

\maketitle




\section{Introduction}
The interaction of elastic waves with dislocations in elastic solids has been studied over several decades, since the pioneering work of Nabarro \cite{nabarro}, Eshelby \cite{eshelby,eshelby2}, Granato and L\"ucke \cite{gl1,gl2} and Mura \cite{mura}. Nabarro \cite{nabarro} and Eshelby \cite{eshelby,eshelby2} noted that the mathematics describing the interaction of screw dislocations with anti-plane elastic waves in a two-dimensional continuum is the same as that of electromagnetic waves in interaction with point charges, also in two dimensions. A translation of the well-studied electrodynamics allowed these researchers to obtain expressions for the elastic radiation emitted by a screw dislocation in arbitrary motion. The more difficult case of the radiation generated by an edge dislocation in arbitrary motion was solved by Mura \cite{mura}, who also found expression for the radiation generated by a dislocation loop of arbitrary shape undergoing also arbitrary motion. The converse problem, the response of a dislocation loop to an incoming, time dependent stress wave, was solved by Lund \cite{lund88} using arguments of energy and momentum conservation.

Granato and L\"ucke \cite{gl1,gl2} formulated a theory for the propagation of an averaged acoustic wave in the presence of many dislocation segments, using a string model for the dislocation pioneered by Kohler \cite{koehler}. Their results have been widely used to interpret experimental results of mechanical damping and modulus change that are both frequency and strain-amplitude dependent. This theory is a scalar theory that cannot distinguish between longitudinal and transverse waves, nor between edge and screw dislocations. Hence, it could not account for results that depend on the polarization of the waves. Also, attempts at using it to  {explain} thermal conductivity  {measurements did not take into account the attenuation due to loss of coherence nor the difffusive behavior of incoherent contributions \cite{glthermal,anderson}}.

In recent years, Maurel, Lund, and collaborators have revisited the theory of the dislocation-wave interaction. Using the equations of \cite{lund88} they obtained explicit expressions for the scattering of an elastic wave by screw and  edge dislocations in two dimensions \cite{MMLjasa1}, and by pinned dislocation segments \cite{singledisloc} and circular dislocation loops \cite{natalia} in three. In addition, and using multiple scattering methods that go back to the work of Foldy \cite{foldy}, Karal and Keller \cite{karalkeller} and Weaver \cite{weaver1}, they obtained expressions for the effective velocity and attenuation of an elastic wave moving through a maze of randomly placed dislocations in two \cite{JF}  and three \cite{manydislocs} dimensions. These results generalized the theory of Granato-L\"ucke to keep track of the wave polarization and vector character of the dislocations. In this way previously unexplained experimental results {concerning the different response of materials to longitudinal and transverse loadings }found an explanation \cite{qlt}. More interestingly, they suggested a way to use ultrasound as a non-intrusive wave to measure dislocation density. This possibility has been recently shown to be feasible, and leads to measurements more accurate than those obtained with X-ray diffraction \cite{actamat,JOM}. The multiple scattering theory of elastic waves has also been studied in relation to propagation in polycrystals \cite{thomson,yang,li} and composites \cite{martin,sumiya}.

Having studied the behavior of coherent waves, the question naturally arises as to the behavior of incoherent waves. Diffusion techniques developed to deal with the localization of de Broglie waves that describe the quantum mechanics of electrons in interaction with randomly placed scatterers have been used to study the behavior of classical waves, a central role being played by the Bethe-Salpeter  (BS) equation \cite{sheng,RMS,ScatRMP}.  The behavior of elastic waves in interaction with a variety of scatterers and within a variety of geometries has been studied by Kirkpatrick \cite{kirkpatrick}, Weaver \cite{weaver}, and Van Tiggelen and collaborators \cite{vantiggelenII,tregoures,tregouresII}. Also, an early approach that uses energy transport equations \cite{papwm} has been widely applied.

The diffusion of elastic waves, albeit in their quantized form, phonons, is central to thermal transport in materials{, a topic of much current concern}. Yet, surprisingly little appears to be quantitatively known about the role played by the interaction of elastic waves with dislocations \cite{thermal-dislocs} in thermal transport. It stands to reason then that a detailed study of the diffusive behavior of elastic waves in interaction with many, randomly placed, dislocations, should be undertaken. Since elastic waves are vector waves and dislocations are linear objects characterized by their tangent and Burgers vectors, the tensor algebra associated with the proposed study appears daunting. A first step should be to consider a simplified setting that captures the essence of the problem. This paper is devoted to precisely this aim: it studies the diffusive behavior, in two dimensions, of an anti-plane elastic wave in interaction with many, randomly placed, screw dislocations.

\subsection{Organization of this paper}
This paper is organized as follows: Section \ref{sec:screw} recalls existing results for the interaction of anti-plane waves with screw dislocations. It includes the behavior of coherent waves when many such dislocations are present, as well as a recent result of Churochkin et al. \cite{RG} concerning the summability to all orders of the perturbation expansion needed to make sense of the theory. Section \ref{BSWT} constructs the Bethe-Salpeter equation and establishes the Ward-Takahashi identity that relates the coherent with the incoherent kernels. Section \ref{lowfreq} solves the BS equation in the low frequency limit needed to obtain a diffusion behavior and an expression for the diffusion coefficient (Eqn. (\ref{DCFinExpl})) is obtained. Section \ref{disc} has a discussion and final comments. A number of computations are carried out in several appendices.

\section{Interaction of anti-plane elastic waves with screw dislocations.}
\label{sec:screw}
The interaction of an anti-plane wave with a single dislocation was studied by Eshelby \cite{eshelby}, Nabarro \cite{nabarro} and by Maurel et al. \cite{MMLjasa1}. The coherent behavior that emerges when an anti-plane wave interacts with many, randomly located screws, was elucidated by Maurel et al \cite{JF}, using the following equation of motion in the frequency domain:
\beq\label{waveanti}
\left(\nabla^{2}+k^{2}_{\beta}\right)v({\vec x},\omega)=-V_{{(0)}}({\vec x},\omega)v({\vec x},\omega)
\eeq
where $v$ is particle velocity as a function of two-dimensional position $\vec x$ and frequency $\om$, $k^{2}_{\beta}=\omega^2/c_T^2$ with $c_T^2 = \mu/\rho$,and the potential $V_{{(0)}}$ is
\beq\label{potanti0}
\left. V_{{(0)}}({\bf x},\omega)=\sum\limits_{n=1}^{N}{\cal A}^{n}_{{(0)}} \;
 \frac{\partial}{\partial x_a}  \delta ( \vec x-\vec X^{n}_0 )\;
{\frac{\partial}{\partial x_a}} \right|_{\vec x=\vec X^{n}_0} \, ,
\eeq
and
 \beq\label{prefanti0}
{\cal A}^{n}_{{(0)}}=\frac{\mu}{M}\left(\frac{b^{n}}{\omega}\right)^{2} \, .
\eeq
Here the superscript $^{n}$ denotes the corresponding characteristics of the  n-th screw dislocation, $\rho$ is the (two dimensional) mass density, $\mu$ is the shear modulus and $M= (\mu b^2/4\pi c_T^2) \ln (\delta/\epsilon)$ is the usual mass per unit length of a screw dislocation with $\delta$ and $\epsilon$ a long- and short- distance cut-off, respectively. {Eqns. (\ref{potanti0},\ref{prefanti0}) were obtained in \cite{MMLjasa1} using an approximation that neglected the Peierls-Nabarro (PN) force \cite{PN}, and led to a scattering cross section that diverges at low frequencies. The origin of this divergence can be understood qualitatively: since the only length scale present in the problem is the wavelength, the scattering cross section will have to be proportional to it, thus diverging as the wavelength grows. The detailed calculation bears out this qualitative reasoning \cite{MMLjasa1}. Since in the present work we wish to explore a diffusion behavior associated with long wavelengths and low frequencies, we shall introduce a Peierls-Nabarro restoring force as well as a viscous damping into the dislocation dynamics, whose equation of motion {, for small oscillations around the PN minimum,} will then be
\beq
\label{eq_screwPN2}
M \ddot X_b  + B \dot X_b + \gamma X_b = \mu b \epsilon_{bc} \frac{\partial u}{\partial x_c} (\vec X(t),t) \, .
\eeq
This dynamics leads to the following equation of motion, that is a generalization of (\ref{waveanti}):
\beq\label{waveanti2}
\left(\nabla^{2}+k^{2}_{\beta}\right)v({\vec x},\omega)=-V({\vec x},\omega)v({\vec x},\omega)
\eeq
with
\beq\label{potanti}
\left. V({\bf x},\omega)=\sum\limits_{n=1}^{N}{\cal A}^{n} \;
 \frac{\partial}{\partial x_a}  \delta ( \vec x-\vec X^{n}_0 )\;
{\frac{\partial}{\partial x_a}} \right|_{\vec x=\vec X^{n}_0}
\eeq
and
 \beq\label{prefanti}
{\cal A}^{n}=\frac{\mu b^2}{M} \frac{1}{\omega^2 {+} \imath \omega (B/M) -\omega^2_0 }
\eeq
where $\omega_0^2 \equiv \gamma /M$.
We shall consider all dislocations have a Burgers vector of the same magnitude, but possibly different sign.

\subsection{Scattering by a single dislocation in the first Born approximation}
Starting from Eqn. (\ref{waveanti2}) and following the reasoning of \cite{MMLjasa1} it is a simple matter to show that  the scattering amplitude for the scattering of an anti plane wave by a screw dislocation located at the origin of coordinates is, in the first Born approximation,
\beq
\label{scatampl}
f(\theta) = -\frac{\mu b^2}{2M} \frac{e^{\imath \pi /4}}{\sqrt{2\pi c_T^3}} \frac{\om^{3/2}}{\om^2 -\om_0^2 {+} \imath \om B/M} \cos \theta
\eeq
from which the total scattering cross section $\sigma^s \equiv \int |f(\theta)|^2 d\theta$ follows:
\beq
\sigma^s = \frac{\mu^2 b^4}{8M^2 c_T^3} \frac{\om^3}{(\om^2 -\om_0^2)^2 + \om^2 (B/M)^2} \, .
\eeq
This cross section vanishes at low frequencies, that is, for $\om \ll \om_0$:
\beq
\sigma_s \sim \frac{\om^3}{\om_0^3} \frac{c_T}{\om_0} \, .
\eeq

\subsection{Coherent behavior}
The coherent behavior of the elastic wave in the presence of many dislocations is described by the average Green function $\langle G \rangle$, where $G$ is the solution of
\beq
\rho\omega^2 G(\vec x,\omega)+\mu\nabla^{2} G(\vec x,\omega)=   - \mu V({\bf x},\omega)G(\vec x,\omega) - \delta(\vec x)  \, .
\label{greeneq}
\eeq
and the brackets denote an average over the random distribution of dislocations.

The average Green function is obtained as the solution of the Dyson equation
\beq
\label{Dyson}
\langle G \rangle^{-1} = (G^0 )^{-1} - \Sigma
\eeq
where
\beq
\label{freegreen}
G^0 =\frac{1}{\mu k^{2}-\rho \omega^{2}}
\eeq
is the free space Green function and  $\Sigma$, the mass (or self-energy) operator, is given by
\beq
\label{MO}
\Sigma = \langle T \rangle - \langle T \rangle G^0 \Sigma
\eeq
with $T$ the T-matrix, given in terms of the ``potential'' $V$ by
\beq
\label{Tmatrix}
T = V + VG^0 T \, .
\eeq
In a perturbation approach this last equation is developed to give
\beq
T = V + VG^0V + VG^0VG^0V + \cdots
\eeq

When the scatterers are statistically independent of each other (ISA, or Independent Scattering Approximation) it is easy to show that
\beq
\Sigma = n \int \langle t \rangle d\vec X_0^n
\label{massISA}
\eeq
where $t$ is the T-matrix for scattering by a single object, as defined in \cite{RG}. Maurel et al. \cite{JF} computed the mass operator to second order in perturbation theory. It is shown in Appendix \ref{Apa} that, due to the point-like nature of the interaction (\ref{potanti}) the perturbation series is geometric and can be summed to all orders to obtain the following expression for the Green function $<G>^{+}({\bf k},\omega)$ for the outgoing waves in the momentum space
\beq
\label{green}
<G>^{+}({\bf k},\omega)=\frac{1}{\rho \omega^{2}\left\{\frac{k^2}{K^{2}}-1\right\}}
\eeq
 {with}
\begin{eqnarray}\label{poles}
K &=& \frac{\omega}{c_{T}\sqrt{\left[1-\frac{\Sigma}{\rho c_{T}^{2}k^{2}}\right]}}= \frac{\omega}{c_{T}\sqrt{1+n{\cal T}} }\\
\Sigma & = & -n\mu {\cal T} k^2 \label{massop} \\
{\cal T} & \equiv & {<\frac{{\mathcal A}^{n}}{1+\frac{\mu{\mathcal A}^{n}I}{4\pi}}> = \frac{\mu b^2}{M} \frac{1}{\om^2 -\om_{0R}^2 + i \left( \om \frac BM + \frac{\mu b^2 \om^2}{8Mc_T^2} \right)}    }
\label{eq:tee}
\end{eqnarray}
 with {$2\mu I = \Lambda^2+ i\pi k_\beta^2$} and $c_{T}^{2}={\mu}/{\rho}${, where $\Lambda$ is a short distance (high wave number) cut-off that is absorbed through a renormalization of the Peierls-Nabarro frequency:
\beq
\om_{0R}^2 \equiv \om_0^2 - \frac{\mu b^2}{8\pi M} \Lambda^2 \, ,
\eeq
 see Appendix \ref{Apa} for a discussion}.
The incoming waves, related to $<G>^{-}({\bf k},\omega)$  {and} $\Sigma^{-}({\bf k},\omega)$, are described by the complex conjugate form of  Eqns. (\ref{green}) and (\ref{massop}).

{Eqn. (\ref{poles}) provides an effective velocity
\beq
v \equiv \frac{\om}{Re[K]} ={\frac{c_{T}|1+n{\cal T}|}{\sqrt{\frac{1+nRe[{\cal T}]+|1+n{\cal T}|}{2}}}}
\eeq
and attenuation length
\beq\label{atl}
l \equiv \frac{1}{2 Im[K]} ={\frac{c_{T}|1+n{\cal T}|}{2\om\sqrt{\frac{|1+n{\cal T}|-\left(1+nRe[{\cal T}]\right)}{2}}}}
\eeq
}

\section{Bethe-Salpeter equation and Ward-Takahashi Identity}
\label{BSWT}

\subsection{BS equation}
  The diffusive transport regime is determined by the two-point correlation  $<G^{+}G^{-}>$ in the low frequency, long wavelength limit. The objective now is to find, in Fourier space, a diffusive pole structure, $<G^{+}G^{-}>\, \sim (\imath \Omega +Dq^2)^{-1}$ and to identify $D$ as a diffusion coefficent ~\cite{sheng}. The first step to be taken is the construction of Bethe-Salpeter (BS) equation. This can be achieved with a reasoning analogous to the one that leads to the Dyson equation: First, the intensity $<G^{+}G^{-}>$ is written as
\begin{eqnarray}\label{Intensity}
<G^{+}G^{-}> &= &<G^{+}><G^{-}>+
\left(<G^{+}G^{-}>-<G^{+}><G^{-}>\right)  \nonumber \\
 &=&  <G^{+}><G^{-}> +
<G^{+}><G^{-}><G^{+}>^{-1}<G^{-}>^{-1} \nonumber \\
& & \times \left(<G^{+}G^{-}>-<G^{+}><G^{-}>\right)
<G^{+}G^{-}>^{-1}<G^{+}G^{-}> \nonumber\\
& =& <G^{+}><G^{-}>
+<G^{+}><G^{-}> \nonumber \\
& & \times \left(<G^{+}>^{-1}<G^{-}>^{-1}-<G^{+}G^{-}>^{-1}\right)
<G^{+}G^{-}> \, .
\end{eqnarray}
Defining the irreducible part $\mathcal{K}$ as
\begin{eqnarray}\label{BSE}
\mathcal{K}=<G^{+}>^{-1}<G^{-}>^{-1}-<G^{+}G^{-}>^{-1} \,
\end{eqnarray}
and substitution into (\ref{Intensity}) gives the BS equation the well known form
\beq\label{IntensityForm}
<G^{+}G^{-}>  =
<G^{+}><G^{-}>
 + <G^{+}><G^{-}>\mathcal{K}<G^{+}G^{-}> \, ,
\eeq
where $\mathcal{K}$ plays a role similar to the role played by the self-energy $\Sigma$ in Dyson's equation (\ref{Dyson}).

It is convenient to work with Eqn. (\ref{IntensityForm}) in Fourier space. Introducing the Fourier transformation as \cite{sheng}
\begin{eqnarray}\label{BSft}
G^{+}(\mathbf{x}_{1},\mathbf{x}^{\prime}_{1};\omega^{+}) & = &\int\limits_{\mathbf{k}_{1}} \int\limits_{\mathbf{k}^{\prime}_{1}}e^{\imath \mathbf{k}_{1}\mathbf{x}_{1}} G^{+}(\mathbf{k}_{1},\mathbf{k}^{\prime}_{1};\omega^{+})e^{-\imath \mathbf{k}^{\prime}_{1}\mathbf{x}^{\prime}_{1}} \nonumber \\
G^{-}(\mathbf{x}_{2},\mathbf{x}^{\prime}_{2};\omega^{-}) & = & \int\limits_{\mathbf{k}_{2}} \int\limits_{\mathbf{k}^{\prime}_{2}}e^{-\imath \mathbf{k}_{2}\mathbf{x}_{2}} G^{-}(\mathbf{k}^{\prime}_{2},\mathbf{k}_{2};\omega^{-})e^{\imath \mathbf{k}^{\prime}_{2}\mathbf{x}^{\prime}_{2}} \, , \nonumber \\
 & &
\end{eqnarray}
we obtain
\beq
\label{BSft2}
<G^{+}G^{-}>=\int\limits_{\mathbf{k}} \int\limits_{\mathbf{k}^{\prime}}\int\limits_{\mathbf{q}}\Phi({\bf k},{\bf k}^{\prime};{\bf q},\Omega)
e^{\imath\left(\mathbf{k}\mathbf{r}-\mathbf{k}^{\prime}\mathbf{r}^{\prime}+\mathbf{q}\left(\mathbf{R}-\mathbf{R}^{\prime}\right)\right)}
\eeq
where
\begin{eqnarray}\label{intensity}
\Phi({\bf k},{\bf k}^{\prime};{\bf q},\Omega)\equiv <G^{+}(\mathbf{k}^{+},\mathbf{k}^{\prime +},\omega^{+})G^{-}(\mathbf{k}^{\prime-},\mathbf{k}^{-},\omega^{-})>
\end{eqnarray}
 {with}
\beq
\mathbf{k}^{\pm}=\mathbf{k}\pm\frac{\mathbf{q}}{2},\quad\omega^{\pm}=\omega\pm\frac{\Omega}{2} \, .\nonumber
\eeq
And space variables have been determined through
\begin{eqnarray}\label{BSspace}
\mathbf{x}_{1}=\mathbf{R}+\frac{\mathbf{r}}{2}, & &
\mathbf{x}_{2}=\mathbf{R}-\frac{\mathbf{r}}{2}\\
\mathbf{x}^{\prime}_{1}=\mathbf{R}^{\prime}+\frac{\mathbf{r}^{\prime}}{2}, & &
\mathbf{x}^{\prime}_{2}=\mathbf{R}^{\prime}-\frac{\mathbf{r}^{\prime}}{2}\nonumber\\
\mathbf{k}_{1}=\mathbf{k}^{+}=\mathbf{k}+\frac{\mathbf{q}}{2}, & &
\mathbf{k}^{\prime}_{1}=\mathbf{k}^{\prime+}=\mathbf{k}^{\prime}+\frac{\mathbf{q}}{2}\nonumber\\
\mathbf{k}^{\prime}_{2}=\mathbf{k}^{\prime-}=\mathbf{k}^{\prime}-\frac{\mathbf{q}}{2}, & & \mathbf{k}_{2}=\mathbf{k}^{-}=\mathbf{k}-\frac{\mathbf{q}}{2} \, .\nonumber
\end{eqnarray}
Applying the inverse Fourier transform \cite{stark}
\begin{eqnarray}\label{BSIft}
\int d\left(\mathbf{R}-\mathbf{R}^{\prime}\right)d\mathbf{r}d\mathbf{r}^{\prime}
e^{-\imath\left(\mathbf{k}\mathbf{r}-\mathbf{k}^{\prime}\mathbf{r}^{\prime}+\mathbf{q}\left(\mathbf{R}-\mathbf{R}^{\prime}\right)\right)} \end{eqnarray}
to Eqn. (\ref{IntensityForm}) the BS equation in momentum space is obtained as
\bea
\label{BSmom}
\Phi({\bf k},{\bf k}^{\prime};{\bf q},\Omega)=<G^{+}><G^{-}>({\bf k};{\bf q},\Omega)\delta_{{\bf k},{\bf k}^{\prime}}+\\
\int\limits_{\mathbf{k}^{\prime\prime}}<G^{+}><G^{-}>({\bf k};{\bf q},\Omega)\mathcal{K}({\bf k},{\bf k}^{\prime\prime};{\bf q},\Omega)\Phi({\bf k}^{\prime\prime},{\bf k}^{\prime};{\bf q},\Omega)\nonumber
\eea
with $\delta_{{\bf k},{\bf k}^{\prime}}=(2\pi)^{2}\delta({\bf k}-{\bf k}^{\prime})$ and, as usual, the integration over the internal momentum variables, i.e. ${\bf k}^{\prime\prime}$, is assumed, with $\int\limits_{\mathbf{k}} \equiv \frac{1}{(2\pi)^{2}}\int d\mathbf{k}$.

To proceed from Eqn. (\ref{BSmom}) to a kinetic form we use the following identity for the  averaged Green's functions:
\beq
\label{BSidentity}
(<G^{+}>^{-1}-<G^{-}>^{-1})<G^{+}><G^{-}>=
<G^{-}>-<G^{+}>
\eeq
Multiplying from the left Eqn. (\ref{BSmom}) by the difference $(<G^{+}>^{-1}-<G^{-}>^{-1})$ and using the property (\ref{BSidentity}) it is straightforward to obtain
\bea
\label{BSkinetic}
(<G^{+}>^{-1}-<G^{-}>^{-1})\Phi& = & (<G^{-}>-<G^{+}>)\\
& &
\times \left(
\delta_{{\bf k},{\bf k}^{\prime}}+
\int\limits_{\mathbf{k}^{\prime\prime}}\mathcal{K}({\bf k},{\bf k}^{\prime\prime};{\bf q},\Omega)\Phi({\bf k}^{\prime\prime},{\bf k}^{\prime};{\bf q},\Omega)\right)\nonumber
\eea
and, using Dyson's equation (\ref{Dyson}) and the explicit form for the free space Green function(\ref{freegreen}) this can be rewritten as
\bea
\label{BSkineticshape}
(\mu \left((k^{+})^{2}-(k^{-})^{2}\right)-\rho \left((\omega^{+})^{2}- (\omega^{-})^{2}\right)+ \Sigma^{-}- \Sigma^{+})\Phi=\\
(<G^{-}>-<G^{+}>)
\left(
\delta_{{\bf k},{\bf k}^{\prime}}+
\int\limits_{\mathbf{k}^{\prime\prime}}\mathcal{K}({\bf k},{\bf k}^{\prime\prime};{\bf q},\Omega)\Phi({\bf k}^{\prime\prime},{\bf k}^{\prime};{\bf q},\Omega)\right)\nonumber \, .
\eea
Introducing the notation
\bea
\label{parameters}
P({\bf k};{\bf q}) & \equiv &\frac{1}{2\imath\rho}\left(L(\mathbf{k}^{-})-L(\mathbf{k}^{+})\right)=\frac{\mu\mathbf{k}\cdot\mathbf{q}}{\imath\rho}\\
\Delta G({\bf k};{\bf q},\Omega) & \equiv & \frac{1}{2\imath\rho}\left(<G>^{-}(\mathbf{k}^{-},\omega^{-})-<G>^{+}(\mathbf{k}^{+},\omega^{+})\right)\nonumber\\ \Delta\Sigma({\bf k};{\bf q},\Omega)  & \equiv &
\frac{1}{2\imath\rho}\left(\Sigma^{-}(\mathbf{k}^{-},\omega^{-})-\Sigma^{+}(\mathbf{k}^{+},\omega^{+})\right)\nonumber\\
L(\mathbf{k}^{\pm}) & = &-\mu k^{\pm}_{l}k^{\pm}_{l}\nonumber \, .
\eea
 yields  the kinetic form of the BS equation
\beq\label{BSfinal}
\left[\imath\omega\Omega+P({\bf k};{\bf q})\right]\Phi({\bf k},{\bf k}^{\prime};{\bf q},\Omega)+
\int\limits_{\bf{k}^{\prime\prime}}U({\bf k},{\bf k}^{\prime\prime};{\bf q},\Omega)\Phi({\bf k}^{\prime\prime},{\bf k}^{\prime};{\bf q},\Omega)
=\delta_{{\bf k},{\bf k}^{\prime}}\Delta G({\bf k};{\bf q},\Omega)
\eeq
with
\beq\label{potBS}
U({\bf k},{\bf k}^{\prime};{\bf q},\Omega) \equiv
\Delta\Sigma({\bf k};{\bf q},\Omega)\delta_{{\bf k},{\bf k}^{\prime}}- \Delta G({\bf k};{\bf q},\Omega)\mathcal{K}({\bf k},{\bf k}^{\prime};{\bf q},\Omega) \, .
\eeq

\subsection{pre-WTI}
The manipulations of the previous subsection are quite general and do not rely on the specifics of the interaction (\ref{potanti}). We now specialize to the case at hand. A Ward-Takahashi identity (WTI) is needed to relate the mass operator $\Sigma$ with the irreducible $\mathcal{K}$. To this end, a preliminary identity (``pre-WTI''), Eqn. (\ref{preWTI}) below, will be obtained.

In order to get the pre-WTI identity we start with the dynamical equation assuming the most general form of the source for the antiplane case $s(\vec x,t)$ \cite{MMLjasa1}:
\begin{equation}\label{waveantiplane}
\rho \frac{\partial^2}{\partial t^2}
v(\vec x,t)-\mu \nabla^{2} v
(\vec x,t)= s(\vec x,t)
\end{equation}
where the source is defined as
\begin{equation}
 s(\vec x,t)=\mu b\epsilon_{ab}\dot{X}_{b}\frac{\partial}{\partial x_{a}}\delta (\vec x-\vec X)
\label{source}
\end{equation}
It yields,
\begin{equation}
 \int\limits_{\vec x}s(\vec x,t)\equiv 0
\label{sourceprop}
\end{equation}
We are requiring that any physically relevant approximations for our source should follow the fundamental property from the Eqn. (\ref{sourceprop}). In particular, the approximation that the dislocation position never deviates significantly from its equilibrium position $\vec X_0$  always brings the source to the Fourier transformed form
\begin{equation}
 s(\vec x,\omega)=\mu V^{n}(\vec x,\omega,\vec X_0)v(\vec x,\omega)
 \label{sourceapp}
\end{equation}
with
\begin{equation}
V^{n}(\vec x,\omega,\vec X_0)=\left. {\cal A}^{n} \;
 \frac{\partial}{\partial x_a}  \delta ( \vec x-\vec X^{n}_0 )\;
{\frac{\partial}{\partial x_a}} \right|_{\vec x=\vec X^{n}_0} \, .
\end{equation}
 Eqns. (\ref{sourceprop},\ref{sourceapp}) lead to an identity for the corresponding Green function $G^{1}(\vec x_{1},\vec x_{1}^{\prime},\omega_{1})$
\begin{eqnarray}
\int\limits_{\vec x_{1}}s(\vec x_{1},\omega_{1})=\int\limits_{\vec x_{1}}\mu V^{n}(\vec x_{1},\omega_{1},\vec X_0)v(\vec x_{1},\omega_{1})\equiv 0\nonumber\\
\Rightarrow \int\limits_{\vec x_{1}}\mu V^{n}(\vec x_{1},\omega_{1},\vec X_0)G^{1}(\vec x_{1},\vec x_{1}^{\prime},\omega_{1})\equiv 0
 \label{identity}
\end{eqnarray}
so that, using Eqn. (\ref{identity}), one can easily prove that
\begin{eqnarray}\label{difference}
\omega^{2}_{2}\int\limits_{\vec x_{2}}\int\limits_{\vec x_{1}}\mu V^{n}(\vec x_{1},\omega_{1},\vec X_0)G^{1}(\vec x_{1},\vec x_{1}^{\prime},\omega_{1})G^{2}(\vec x_{2},\vec x_{2}^{\prime},\omega_{2}) & & \nonumber\\
-\omega^{2}_{1}\int\limits_{\vec x_{2}}\int\limits_{\vec x_{1}}\mu V^{n}(\vec x_{2},\omega_{2},\vec X_0)G^{2}(\vec x_{2},\vec x_{2}^{\prime},\omega_{2})G^{1}(\vec x_{1},\vec x_{1}^{\prime},\omega_{1})
& \equiv & 0 \, .
\end{eqnarray}
This last identity enables the  construction of a pre-WTI following Refs. \cite{barabanenkovI,barabanenkovII}.
This is achieved as follows:
a system of equations is constructed, writing out Eqn. (\ref{greeneq}) twice for $G({\bf 1})=G(\vec x_{1},\vec x^{\prime}_{1},\omega_{1})$ and $G({\bf 2})=G(\vec x_{2},\vec x^{\prime}_{2},\omega_{2})$ respectively. Next, the first and second equations of the resulting system are multiplied by $\omega_{2}^{2}G(\mathbf{2})$ and $\omega_{1}^{2}G(\mathbf{1})$ respectively, and substracted from each other. Using the property (\ref{difference}) of the potential and integrating over $\vec x_{1}$ and $\vec x_{2}$ we come to the following identity, that does not explicitly  involve the potential $\mu V$:
\beq\label{preWTIint}
\int\limits_{\textbf{x}_{1}}\int\limits_{\textbf{x}_{2}}\left(\left(\omega_{2}^{2}L({\bf 1})-\omega_{1}^{2}L({\bf 2})\right)\left(G(\mathbf{1})G(\mathbf{2})\right)+G(\mathbf{2})\delta(\textbf{x}_{1}-\textbf{x}^{\prime}_{1})\omega_{2}^{2}
-G(\mathbf{1})\delta(\textbf{x}_{2}-\textbf{x}^{\prime}_{2})\omega_{1}^{2}\right)\equiv 0
\eeq
Here, $\int\limits_{\textbf{x}}=\int d\textbf{x}$, $L=L(\imath \frac{\partial}{\partial \textbf{x}})$.
The identity in Eqn. (\ref{preWTIint}) must be fulfilled for arbitrary values of the external parameters such as $\textbf{x}^{\prime}_{1,2}$ and $\omega_{1,2}^{2}$. This is possible only if the expression in the brackets is equal to zero. Its subsequent averaging gives the pre-WTI:
\beq\label{preWTI}
\lim\limits_{{\substack{\textbf{x}_{1}\rightarrow\textbf{x}_{2}\\ \textbf{x}^{\prime}_{1}\rightarrow\textbf{x}^{\prime}_{2}}}}\left(\left(g_{2}L({\bf 1})-g_{1}L({\bf 2})\right)<G(\mathbf{1})G(\mathbf{2})> \equiv <G(\mathbf{1})\delta(\textbf{x}_{2}-\textbf{x}^{\prime}_{2})g_{1}-G(\mathbf{2})\delta(\textbf{x}_{1}-\textbf{x}^{\prime}_{1})g_{2}>\right)
\eeq
with $g_{1,2}=g(\omega_{1,2})=\omega_{1,2}^{2}$.

\subsection{WTI}
The WTI relates the vertex $\cal{K}$ and the self energy $\Sigma$, using  Eqn. (\ref{preWTI}). To obtain this relationship we modify the BS Eqn. (\ref{BSmom}), multiplying it from the right  with $\left(g_{2}L(\mathbf{k}^{\prime +})-g_{1}L(\mathbf{k}^{\prime -})\right)$
and integrating over $\mathbf{k}^{\prime}$. This gives
\begin{eqnarray}\label{BSmod}
\int\limits_{\mathbf{k}^{\prime}}\Phi({\bf k},{\bf k}^{\prime};{\bf q},\Omega)\left(g_{2}L(\mathbf{k}^{\prime +})-g_{1}L(\mathbf{k}^{\prime -})\right) & = & \\ & & \hspace{-20em} \int\limits_{\mathbf{k}^{\prime}}\left(<G^{+}><G^{-}>({\bf k};{\bf q},\Omega)\delta_{{\bf k},{\bf k}^{\prime}}
+  \int\limits_{\mathbf{k}^{\prime\prime}}<G^{+}><G^{-}>({\bf k};{\bf q},\Omega)\mathcal{K}({\bf k},{\bf k}^{\prime\prime};{\bf q},\Omega)\Phi({\bf k}^{\prime\prime},{\bf k}^{\prime};{\bf q},\Omega)\right) \nonumber \\
&& \times \left(g_{2}L(\mathbf{k}^{\prime +})-g_{1}L(\mathbf{k}^{\prime -})\right)\nonumber
\end{eqnarray}
Using the Fourier-transformed Eqn. (\ref{preWTI}) with arguments $\mathbf{1}$ (resp  $\mathbf{2}$) replaced by ``$+$'' (resp ``$-$'') in  Eqn.(\ref{BSmod}) and then multiplying the obtained expression with $<G^{+}>^{-1}<G^{-}>^{-1}({\bf k};{\bf q},\Omega)$ from the left we come to the result
\begin{eqnarray}\label{WTIres}
<G^{-}>^{-1}({\bf k};{\bf q},\Omega)g_{1}-<G^{+}>^{-1}({\bf k};{\bf q},\Omega)g_{2}
& = & \left(g_{2}L(\mathbf{k}^{+})-g_{1}L(\mathbf{k}^{-})\right)\\
& & \hspace{-15em} + \int\limits_{\mathbf{k}^{\prime\prime}}\left(\mathcal{K}({\bf k},{\bf k}^{\prime\prime};{\bf q},\Omega)\right)\left(<G^{+}>({\bf k}^{\prime\prime};{\bf q},\Omega)g_{1}-<G^{-}>({\bf k}^{\prime\prime};{\bf q},\Omega)g_{2}\right)\nonumber \, .
\end{eqnarray}
Introducing the explicit form of the averaged Green function, Eqns. (\ref{Dyson},\ref{freegreen}), the following WTI, relating $\Sigma$ and $\mathcal{K}$, is obtained:
\begin{eqnarray}\label{WTIsim}
\Sigma^{+}({\bf k};{\bf q},\Omega))g_{2}-\Sigma^{-}({\bf k};{\bf q},\Omega)g_{1}
& = & \\
&& \hspace{-10em} \int\limits_{\mathbf{k}^{\prime\prime}}\left(\mathcal{K}({\bf k},{\bf k}^{\prime\prime};{\bf q},\Omega)\right)\left(<G^{+}>({\bf k}^{\prime\prime};{\bf q},\Omega)g_{1}-<G^{-}>({\bf k}^{\prime\prime};{\bf q},\Omega)g_{2}\right)\nonumber
\end{eqnarray}
It is convenient to rewrite this form of the WTI, Eqn. (\ref{WTIsim}), in a different way. By presenting $g_{1}$ and $g_{2}$ as
\begin{eqnarray}\label{wtiaux}
g_{1}=\frac{g_{1}+g_{2}}{2}+\frac{(g_{1}-g_{2})}{2},\quad
g_{2}=\frac{g_{1}+g_{2}}{2}-\frac{(g_{1}-g_{2})}{2}
\end{eqnarray}
and substituting them into Eqn. (\ref{WTIsim}), the commonly used appearance for the WTI is obtained:
\begin{eqnarray}\label{WTImod}
 \left(\Sigma^{+}({\bf k};{\bf q},\Omega)-\Sigma^{-}({\bf k};{\bf q},\Omega)\right)-\int\limits_{\mathbf{k}^{\prime\prime}}\mathcal{K}({\bf k},{\bf k}^{\prime\prime};{\bf q},\Omega)\left(<G^{+}>({\bf k}^{\prime\prime};{\bf q},\Omega)-<G^{-}>({\bf k}^{\prime\prime};{\bf q},\Omega)\right) = \hspace{3em} && \\
\frac{(g_{1}-g_{2})}{g_{1}+g_{2}} \left(\left(\Sigma^{+}({\bf k};{\bf q},\Omega)+\Sigma^{-}({\bf k};{\bf q},\Omega)\right)+\int\limits_{\mathbf{k}^{\prime\prime}}\mathcal{K}({\bf k},{\bf k}^{\prime\prime};{\bf q},\Omega)\left(<G^{+}>({\bf k}^{\prime\prime};{\bf q},\Omega)+<G^{-}>({\bf k}^{\prime\prime};{\bf q},\Omega)\right)\right) && \nonumber
\end{eqnarray}
Eqn. (\ref{WTImod}) can be rewritten in a more compact form, as
\beq\label{WLWTI}
\int\limits_{\mathbf{k}^{\prime\prime}}U(\mathbf{k}^{\prime\prime},\mathbf{k}^{\prime};\mathbf{q},\Omega)= \frac{i}{2}A(\mathbf{k}^{\prime};\mathbf{q},\Omega)\left(g_{+}-g_{-}\right)
\eeq
with $g_{1,2}=g_{\pm}=\omega^{2}_{\pm}$,
\begin{eqnarray}\label{WLUtensor}
U({\bf k},{\bf k}^{\prime};{\bf q},\Omega)&=& \Delta\Sigma({\bf k};{\bf q},\Omega)\delta_{{\bf k},{\bf k}^{\prime}}-\Delta G({\bf k};{\bf q},\Omega)\mathcal{K}({\bf k},{\bf k}^{\prime};{\bf q},\Omega) \, ,\\
 A(\mathbf{k}^{\prime};\mathbf{q},\Omega)&=&\frac{2}{g_{+}+g_{-}}
\left({\cal R}\Sigma(\mathbf{k}^{\prime};\mathbf{q},\Omega)+\int\limits_{\mathbf{k}^{\prime\prime}}{\cal R}G(\mathbf{k}^{\prime\prime};\mathbf{q},\Omega)\mathcal{K}({\bf k^{\prime\prime}},{\bf k^{\prime}};{\bf q},\Omega)\right) \, , \nonumber\\
{\cal R}\Sigma(\mathbf{k}^{\prime};\mathbf{q},\Omega)&=&
\frac{1}{2\rho}\left(\Sigma^{-}(\mathbf{k}^{\prime-},\omega^{-})+\Sigma^{+}(\mathbf{k}^{\prime+},\omega^{+}) \right) \, , \nonumber
\end{eqnarray}
and the operation ${\cal R}$, here defined for the self-energy $\Sigma$, acts in the same way on the Green function $G$.  {It is remarkable that the essential point that enables the derivation of the WTI is the property (\ref{sourceprop}) of the source (\ref{source}). That is, a Peach-Koehler force that acts at the equilibrium position $\vec X_0$ of the dislocation, that in turns perform conservative (i.e., glide, not climb) motion. The specific dislocation dynamics embodied in (\ref{eq_screwPN2}) does not play a role. It does, however, become relevant when we take the long wavelength, low frequency and low density limit, to which we now turn our attention.}

{
\subsection{WTI in the low-density-of-dislocations, low frequency, and long wavelength approximations}
The plan is to solve the BS equation (\ref{BSfinal}) to obtain the two point correlation in Fourier space, $\Phi({\bf k},{\bf k}^{\prime};{\bf q},\Omega)$, in the low frequency ($\Omega \rightarrow 0$) and long wavelength (${\bf q} \rightarrow 0$) limit. Hopefully a diffusion behavior will result. In order to do this we need information about the self energy $\Sigma$ and vertex $\mathcal{K}$. The mass operator $\Sigma$ is given by (\ref{massISA}) in the ISA approximation: scatterers are uniformly distributed through space with uncorrelated positions. This is valid for low scatterer density $n$ and, indeed, (\ref{massISA}) is the leading order term in a low-$n$ approximation scheme. Similarly keeping the lowest order term in dislocation density $n$ for the kernel $\mathcal{K}$ leads to its Boltzmann approximation $\mathcal{K} \rightarrow \mathcal{K}^B$, with
\bea
\mathcal{K}^B({\bf k},{\bf k}^{\prime};{\bf q},\Omega) = n \langle \, t_{\om^+} ({\bf k}^+,{\bf k}^{\prime +}) t_{\om^-}^* ({\bf k}^{\prime -} , {\bf k}^-) \, \rangle
\label{isabol}
\eea
where the frequency dependence of the $T$ matrix has been made explicit \cite{RG}. In order to obtain the diffusive behaviour of our system in this approximation, we need the low frequency and long wavelength asymptotics: the limit $\mathbf{q}\rightarrow0$, $\Omega\rightarrow0$.
It can be easily seen from Eqns. (\ref{WLWTI}) and (\ref{WLUtensor}) that, in this limit, the WTI prescribes the following constraint {for the corresponding part of the kernel} $\mathcal{K}^{B}({\bf p}^{\prime\prime},{\bf k}^{\prime};{\bf 0},0)$:
\beq
\label{WLWTIzero}
\Delta \Sigma(\mathbf{k}^{\prime})=\int\limits_{\mathbf{p}^{\prime\prime}}\Delta G(\mathbf{p}^{\prime\prime})\mathcal{K}^{B}({\bf p}^{\prime\prime},{\bf k}^{\prime};{\bf 0},0) = \int\limits_{\mathbf{p}^{\prime\prime}}\mathcal{K}^{B}({\bf k}^{\prime},{\bf p}^{\prime\prime};{\bf 0},0)\Delta G(\mathbf{p}^{\prime\prime})
\eeq
The second equality in Eqn. (\ref{WLWTIzero}) is due to the reciprocity of the $K^{B}({\bf p}^{\prime\prime},{\bf k}^{\prime};{\bf Q},\Omega)$, a consequence of the symmetries of the Green's function $G$. It is shown in Appendix \ref{Apb} that this relation holds within the ISA, {in the abscence of viscosity in the dislocation dynamics; i.e., when $B=0$ in (\ref{eq_screwPN2}).}
 }
\section{Low frequency asymptotics and diffusion behavior}
\label{lowfreq}
The following relations for the self energy and for the Green function will prove useful:
\bea\label{SEDiff}
\Delta\Sigma({\bf k};{\bf 0},0) & = & \frac{1}{2\imath\rho}\left(\Sigma^{*}({\bf k})-\Sigma({\bf k})\right)\\
 & = & \frac{\rho c^{2}_{T}k^{2}}{2\imath\rho}\left(\left(1-\frac{\omega^{2}}{(K^{2})^{*}c^{2}_{T}}\right)-\left(1-\frac{\omega^{2}}{K^{2}c^{2}_{T}}\right)\right) \nonumber\\
& = & \frac{-Im[K^{2}]\omega^{2} k^{2}}{K^{2}(K^{2})^{*}} \, .\nonumber \\
\label{GFDiff}
\Delta G({\bf k};{\bf 0},0) & = & \frac{1}{2\imath\rho}\left(<G>^{*}({\bf k})-<G>({\bf k})\right)\\
 & = & \frac{1}{2\imath\rho^{2}\omega^{2}}\left(\frac{(K^{2})^{*}}{k^{2}-(K^{2})^{*}}-\frac{K^{2}}{k^{2}-K^{2}}\right)\nonumber\\
 & = & \frac{-k^{2}}{\rho^{2}\omega^{2}}\frac{Im[K^{2}]}{\left(k^{2}-(K^{2})^{*}\right)\left(k^{2}-K^{2}\right)} \nonumber \\
 & \approx & \frac{-\pi k^{2}}{\rho^{2}\omega^{2}}\delta\left(k^{2}-Re[K^{2}]\right)
 \label{deltaGdelta}
\eea
where the last approximate equality holds in the limit $| Im[K^{2}] | \ll | k^{2}-Re[K^{2}] |$. {The meaning of this inequality in terms of the dislocation parameters is explored in  Appendix \ref{APK}.}

\subsection{Perturbation approach to BS eigenvalue problem}
In this section we follow the approach that was used in Refs.
 \cite{stark,barabanenkovI,barabanenkovII,berman} to study the diffusion of  electromagnetic and acoustic waves: The BS equation written in the form (\ref{BSfinal}), supplemented by {the relation between mass operator $\Sigma$ and kernel $\mathcal{K}$ implemented by the WTI (\ref{WLWTI})},} is solved for the intensity $\Phi$, defined by (\ref{intensity}), in the diffusive limit.  To this end, the BS equation (\ref{BSfinal}) is written in operator form:
 \beq\label{BSH}
\int\limits_{\bf{k}^{\prime\prime}}H({\bf k},{\bf k}^{\prime\prime};{\bf q},\Omega)\Phi({\bf k}^{\prime\prime},{\bf k}^{\prime};{\bf q},\Omega)=\Delta G({\bf k};{\bf q},\Omega)\delta_{{\bf k},{\bf k}^{\prime}} \, .
\eeq
with the operator $H$ defined by
\beq\label{Hoperator}
H\equiv \left[\imath\omega\Omega+P({\bf k};{\bf q})\right]\delta_{\mathbf{k}\mathbf{k}^{\prime\prime}}+U({\bf k},{\bf k}^{\prime\prime};{\bf q},\Omega) \, .
\eeq
It is easy to see, using the explicit form of $U$, and the reciprocity of the tensor $\mathcal{K}$, that $H$ has the following symmetry:
 \beq\label{symmetry}
H({\bf k},{\bf k}^{\prime\prime};{\bf q},\Omega)\Delta G({\bf k}^{\prime\prime};{\bf q},\Omega)=H({\bf k}^{\prime\prime},{\bf k};{\bf q},\Omega)\Delta G({\bf k};{\bf q},\Omega) \, .
\eeq
The solution of the BS equation will be found in terms of the eigenvalues and eigenfunctions of the operator $H$.

The eigenvalue problem for $H$ is set up as follows:
\beq
\label{Hom}
\int\limits_{\bf{k}^{\prime\prime}}H({\bf k},{\bf k}^{\prime\prime};{\bf q},\Omega)f^{\mathbf{r}n}_{kl}({\bf k}^{\prime\prime};{\bf q},\Omega)=
\lambda_{n}\left(\mathbf{q},\Omega\right)f^{\mathbf{r}n}({\bf k};{\bf q},\Omega)
\eeq
where $f^{\mathbf{r}n}({\bf k}^{\prime\prime};{\bf q},\Omega)$  (resp. $f^{\mathbf{l}n}({\bf k}^{\prime\prime};{\bf q},\Omega)$) are right (resp. left) eigenfunctions and $\lambda_{n}\left(\mathbf{q},\Omega\right)$ the corresponding eigenvalues. Following \cite{stark,barabanenkovI,barabanenkovII,berman} we shall assume that the eigenfunctions in (\ref{Hom}) satisfy completeness and orthogonality conditions:
\begin{eqnarray}\label{basis}
\int\limits_{\mathbf{k}}f^{\mathbf{r}m}({\bf k};{\bf q},\Omega)f^{\mathbf{l}n}({\bf k};{\bf q},\Omega)&=& \delta_{mn}\, ,\\
\sum\limits_{n}f^{\mathbf{r}n}({\bf k};{\bf q},\Omega)f^{\mathbf{l}n}({\bf k}^{\prime};{\bf q},\Omega)&=&\delta_{{\bf k}{\bf k}^{\prime}}\nonumber \, .
\end{eqnarray}
Furthermore, the symmetry restriction for the operator $H$ from Eqn. (\ref{symmetry}) determines the relation between left and right eigenfunctions:
\beq
\label{lr}
f^{\mathbf{r}n}({\bf k};{\bf q},\Omega)=\Delta G({\bf k};{\bf q},\Omega)
f^{\mathbf{l}n}({\bf k};{\bf q},\Omega) \, .
\eeq

The eigenfunction properties  (\ref{basis},\ref{lr}) allow for a representation of the solution $\Phi$ of (\ref{BSH})  as a series over the states $n$  \cite{stark,barabanenkovI,barabanenkovII,berman} :
\begin{eqnarray}\label{solution}
\Phi=\sum\limits_{n}\frac{f^{\mathbf{r}n}({\bf k};{\bf q},\Omega)f^{\mathbf{r}n}({\bf k}^{\prime};{\bf q},\Omega)}{\lambda_{n}\left(\mathbf{q},\Omega\right)}
\end{eqnarray}
The existence of a diffusion regime assumes that in the limit $\mathbf{q}\rightarrow0$, $\Omega\rightarrow0$ the intensity $\Phi$ has a pole structure given by the lowest eigenvalue asymptotics $\lambda_{0}\left(\mathbf{q}\rightarrow 0,\Omega\rightarrow 0\right)\rightarrow 0$ that is separated from a regular part\cite{barabanenkovI,barabanenkovII,berman}. Therefore, the whole problem is reduced to the determination of coefficients of a perturbative expansion for $\lambda_{0}\left(\mathbf{q},\Omega\right)$ to second order in $\mathbf{q}$ and first order in $\Omega$ around the point $\mathbf{q}=0$, $\Omega=0$. To do this, Eqn. (\ref{Hom}) has to be treated perturbatively, with the condition that Eqns. (\ref{WLWTI},\ref{symmetry}) hold at every order of the perturbation in $\mathbf{q}$, and $\Omega$ \cite{barabanenkovI,barabanenkovII,berman}.

In order to solve Eq. (\ref{Hom}), we write, in a small $\Omega$ and a small $\mathbf{q}$ approximation,
\begin{eqnarray}\label{Pseries}
H({\bf k},{\bf k}^{\prime\prime};{\bf q},\Omega) & = &
 H({\bf k},{\bf k}^{\prime\prime};{\bf 0},0)+H^{1\Omega}({\bf k},{\bf k}^{\prime\prime};{\bf 0},\Omega) \nonumber \\
& &  +H^{1\mathbf{q}}({\bf k},{\bf k}^{\prime\prime};{\bf q},0)+H^{2\mathbf{q}}({\bf k},{\bf k}^{\prime\prime};{\bf q},0) + \dots \, , \nonumber \\
& & \nonumber \\
\label{Pseries2}
f^{\mathbf{r}0}({\bf k}^{\prime\prime};{\bf q},\Omega) & = &
f({\bf k}^{\prime\prime};{\bf 0},0)+f^{1\Omega}({\bf k}^{\prime\prime};{\bf 0},\Omega) \\
& & +f^{1\mathbf{q}}({\bf k}^{\prime\prime};{\bf q},0)+f^{2\mathbf{q}}({\bf k}^{\prime\prime};{\bf q},0) + \dots \nonumber \\
& & \nonumber \\
\lambda_{0}\left(\mathbf{q},\Omega\right) & = &
\lambda^{1\Omega}\left(\mathbf{0},\Omega\right)+\lambda^{1\mathbf{q}}\left(\mathbf{q},0\right)+\lambda^{2\mathbf{q}}\left(\mathbf{q},0\right) + \dots \nonumber
\label{lambda}
\end{eqnarray}
It is shown in Appendix \ref{pert} that substitution of the above expansions into Eqn. (\ref{Hom}) leads to the following set of coupled integral equations:
\begin{eqnarray}\label{system0}
\int\limits_{\bf{k}^{\prime\prime}}H({\bf k},{\bf k}^{\prime\prime})f({\bf k}^{\prime\prime})& = & 0 \\
\label{Omega}
\int\limits_{\bf{k}^{\prime\prime}}\left(H({\bf k},{\bf k}^{\prime\prime})f^{1\Omega}({\bf k}^{\prime\prime})
+  H^{1\Omega}({\bf k},{\bf k}^{\prime\prime})f({\bf k}^{\prime\prime})\right) & = &
\lambda^{1\Omega}f({\bf k}) \\
\label{system1q}
\int\limits_{\bf{k}^{\prime\prime}}\left(H({\bf k},{\bf k}^{\prime\prime})f^{1\mathbf{q}}({\bf k}^{\prime\prime}) +H^{1\mathbf{q}}({\bf k},{\bf k}^{\prime\prime})f({\bf k}^{\prime\prime})\right) & = & 0 \\
\label{system2q}
\int\limits_{\bf{k}}B P({\bf k};{\bf q})f^{1\mathbf{q}}({\bf k}) & = &
\lambda^{2\mathbf{q}}
\end{eqnarray}
Within Eqns. (\ref{system0}-\ref{system2q}) we have used a shorthand notation for the quantities appearing in Eqn. (\ref{Pseries}), in which the arguments $\mathbf{q}$ and $\Omega$ are omitted. The superscript indicates the variable and the order of perturbation  for the operator $H$, right eigenfunction $f^{\mathbf{r}0}$, and eigenvalue $\lambda_{0}$. The disappearance in Eqns. (\ref{system0}-\ref{system2q}) of contributions from some terms that appear in Eqn. (\ref{Pseries}) is due to the implementation of symmetry and conservation restrictions coming from Eqns. (\ref{WLWTI}) and (\ref{symmetry}),  {as shown in Appendix \ref{pert}. Importantly, the first-order-in-wavenumber contribution to the eigenvalue vanishes: $\lambda^{1\mathbf{q}} = 0$.} Without this result there would be no diffusion behavior.

With the aid of Eqns. (\ref{WLWTI}) and (\ref{system0}), the eigenfunction $f^{\mathbf{r}0}$ at $\mathbf{q}=0$, $\Omega=0$ can be found at once:
\beq\label{zeroorder}
f({\bf k}^{\prime\prime})=B\Delta G({\bf k}^{\prime\prime};{\bf 0},0)
\eeq
with
\beq
B=\frac{1}{\sqrt{\int\limits_{\bf{v}}\Delta G({\bf v};{\bf 0},0)}} .
\label{eq:B}
\eeq
{Note that $\Delta G({\bf v};{\bf 0},0)$ is negative, but only $B^{2}$ appears in the expression for the diffusion constant.}
Integrating Eqn. (\ref{Omega}) over ${\bf k}$ along with the subsequent implementation of the WTI from Eqn. (\ref{WLWTI}) at the corresponding order, leads to the expression for the eigenvalue $\lambda^{1\Omega}$
\begin{eqnarray}\label{firstomega}
\imath\omega\Omega\left(1+a\right)=
\lambda^{1\Omega}
\end{eqnarray}
where we have introduced a parameter $a$ defined by
\begin{eqnarray}
\label{aparameter}
a=\frac{1}{\int\limits_{\bf{k}}f({\bf k})}\int\limits_{\bf{k}^{\prime\prime}} A(\mathbf{k}^{\prime\prime};\mathbf{0},0)f({\bf k}^{\prime\prime}) \, .
\end{eqnarray}
This is an analogue of the well-known parameter that appears in the diffussion of light, which, being positive, renormalizes the phase velocity downwards {to a lower value for a transport velocity}
\cite{barabanenkovI,barabanenkovII,Tiggelen,livdan}. To see that our $a$ is indeed positive, replace  Eqns. (\ref{WLUtensor}) and (\ref{zeroorder}) into Eqn. (\ref{aparameter}) to obtain
\beq
\label{aparaprox}
a  =
\frac{-1}{\rho^{2}\omega^{2}\int\limits_{\bf{v}}\Delta G({\bf v};{\bf 0},0)}\int\limits_{\bf{k}}
Im[\Sigma^{+}(\mathbf{k})G^{+}({\bf k})]\approx
\left(\frac{c_{T}^{2}Re[K^{2}]}{\omega^{2}}-1\right)>0
\eeq
where the last approximate equality is obtained using expression (\ref{FReg}) for the function $F$ defined by Eqn. (\ref{Ffunc}) that enters the integrand of Eqn.(\ref{aparaprox}) as a consequence of the calculation of $Im[\Sigma^{+}(\mathbf{k})G^{+}({\bf k})]$.

\subsection{Explicit form of the diffusion constant}
Within the spectral approach that we are using, Eqns. (\ref{solution}-\ref{system2q}) lead to the following expression for the singular part of the intensity, $\Phi^{sing}$:
\beq\label{sing}
\Phi^{sing}=\frac{f^{\mathbf{r0}}({\bf k};{\bf q},\Omega)f^{\mathbf{r0}}({\bf k}^{\prime};{\bf q},\Omega)}{\lambda^{1\Omega}+\lambda^{2\mathbf{q}}}=\frac{f^{\mathbf{r0}}({\bf k};{\bf q},\Omega)f^{\mathbf{r0}}({\bf k}^{\prime};{\bf q},\Omega)}{\frac{\lambda^{1\Omega}}{-\imath\Omega}\left(-\imath\Omega+\frac{-\imath\Omega\lambda^{2\mathbf{q}}}{\lambda^{1\Omega}q^{2}}q^{2}\right)} \, .
\eeq
Then, with the assistance of Eqns. (\ref{firstomega},\ref{sing}) the diffusion constant can be identified as
\begin{eqnarray}\label{Dconstant}
D & \equiv &-\frac{\imath\Omega\lambda^{2\mathbf{q}}}{q^2\lambda^{1\Omega}}=-\frac{B}{q^{2}\omega\left(1+a\right)}
\int\limits_{\bf{k}}P({\bf k};{\bf q})f^{1\mathbf{q}}({\bf k}) \\
 & = &
\frac{B^{2}}{q^{2}\omega\left(1+a\right)}
\int\limits_{\bf{k}}P({\bf k};{\bf q})
\left(\int\limits_{{\bf k}_{2}}\Phi({\bf k},{\bf k}_{2})P(\mathbf{q};{\bf k}_{2})-\Delta G^{1\mathbf{q}}({\bf k})\right) \label{Dconstant2} \\
& \equiv &  D^{\mathcal{R}} + D_{\Delta G^{1\mathbf{q}}}
\label{Dconstant3}
\end{eqnarray}
with $B$ given by (\ref{eq:B}). To obtain Eqn. (\ref{Dconstant2}), in which the diffusion constant is written as the sum of two terms, we have substituted  the values for $\lambda^{2\mathbf{q}}$, $\lambda^{1\Omega}$ given by Eqns. (\ref{system2q}) and (\ref{firstomega}). The last one ensued from the form of $f^{1\mathbf{q}}({\bf k})$ ({See Appendix \ref{solu}}).
Thus, the expression for the diffusion constant in Eqn. (\ref{Dconstant}) is the sum of two contributions, as defined in (\ref{Dconstant3}).

While the second term in (\ref{Dconstant3}), $D_{\Delta G^{1\mathbf{q}}}$, can be calculated straightforwardly (See Appendix \ref{esti}), and it vanishes, $D_{\Delta G^{1\mathbf{q}}} = 0$, the calculation of the  first term, $D^{\mathcal{R}}$, is more laborious. {Indeed, it depends on the unknown function $\Phi$. However, it is not the complete function that is needed, but an integrated form  over one of its variables that, as we now show, can be expressed as a function of the mass operator $\Sigma$ and the kernel $\mathcal{K}$ using the BS equation and the WTI.} In order to {do this}  we apply, inspired by the treatment of light diffusion  \cite{barabanenkovIV}, the method that uses an auxiliary function $\Psi_{,s}({\bf k})$ defined by the relation
\beq\label{DconstantAUX}
\Psi_{,s}({\bf k})q_{s} \equiv
\int\limits_{{\bf k}^{\prime}}\Phi({\bf k},{\bf k}^{\prime})P(\mathbf{q};{\bf k}^{\prime})=
-\int\limits_{{\bf k}^{\prime}}\Phi({\bf k},{\bf k}^{\prime})\frac{1}{2\imath\rho}\frac{\partial L({\bf k}^{\prime})}{\partial k_{s}}q_{s} \, .
\eeq
Then, from Eqns. (\ref{BSfinal}-\ref{potBS}) the following expression for $\Psi_{,s}({\bf k})$
immediately follows:
\beq\label{DconstantAUXEQ}
P({\bf p})\Psi_{,s}({\bf p})+\Delta\Sigma({\bf p})\Psi_{,s}({\bf p})
-\int\limits_{\bf{p}^{\prime\prime}}
\Delta G({\bf p})\mathcal{K}({\bf p},{\bf p}^{\prime\prime})\Psi_{,s}({\bf p}^{\prime\prime})=
-\Delta G({\bf p})\frac{1}{2\imath\rho}\frac{\partial L({\bf p})}{\partial p_{s}}
\eeq
Eqn. (\ref{DconstantAUXEQ}) can be substantially simplified if we recall the explicit form of the free medium Green's function, Eq.(\ref{freegreen}), along with relations from Eqns. (\ref{Dyson},\ref{parameters}) \cite{JF}. Then,
\beq\label{greenexpl}
\Delta G({\bf p})=\frac{-1}{2i\rho}\left(G^{*}({\bf p})^{-1}-G({\bf p})^{-1}\right)G({\bf p})G^{*}({\bf p})=\left(P({\bf p})+\Delta\Sigma({\bf p})\right)G({\bf p})G^{*}({\bf p})
\eeq
As a next step, we define an angular entity $\Upsilon$ that {is analogous to the coefficient relating transport mean free path and  extinction length in the diffusion of electromagnetic waves \cite{shengIII,KvTRMP}:}
\beq\label{angular}
\Psi_{,s}({\bf p})q_{s}=G({\bf p})G^{*}({\bf p})\Upsilon({\bf p},{\bf q}) \, ,
\eeq
and an integral equation for $\Upsilon$ follows straightforwardly from Eqns. (\ref{DconstantAUXEQ}-\ref{angular}):
\beq\label{angulareq}
\Upsilon({\bf p},{\bf q})-
\int\limits_{\bf{p}^{\prime\prime}}
\mathcal{K}({\bf p},{\bf p}^{\prime\prime})G({\bf p}^{\prime\prime})G^{*}({\bf p}^{\prime\prime})\Upsilon({\bf p}^{\prime\prime},{\bf q})=P({\bf p};{\bf q}) \, .
\eeq
Then, using Eqns. (\ref{DconstantAUX}), (\ref{angular}) and (\ref{Dqeval}),  the diffusion constant from Eq. (\ref{Dconstant}) can be written as
\beq\label{difften}
D =
\frac{B^{2}}{q^{2}\omega\left(1+a\right)}
\int\limits_{\bf{k}}P({\bf k};{\bf q})
\left(G({\bf k})G^{*}({\bf k})\Upsilon({\bf k},{\bf q})\right)
\eeq
Furthermore, guided by the definition from Eqn. (\ref{angular}) we can make a conjecture that tensor $\Upsilon({\bf p},{\bf q})$ should be linear in ${\bf q}$. Keeping in mind this property of $\Upsilon({\bf p},{\bf q})$ we seek for the corresponding solution in the form
 \beq\label{angulsol}
\Upsilon({\bf p},{\bf q})=\alpha P({\bf p};{\bf q})
\eeq
 Then,  by multiplying Eqn. (\ref{angulareq}) with  $P({\bf p};{\bf q})G({\bf p})G^{*}({\bf p})$ from the left and integrating over the ${\bf p}$  we remain with the relation
 \bea\label{angularproof}
\alpha=\left(1-
\frac{\int\limits_{\bf{p}}\int\limits_{\bf{p}^{\prime\prime}}P({\bf p};{\bf q})G({\bf p})G^{*}({\bf p})
\mathcal{K}({\bf p},{\bf p}^{\prime\prime})G({\bf p}^{\prime\prime})G^{*}({\bf p}^{\prime\prime})P({\bf p}^{\prime\prime},{\bf q})}{\int\limits_{\bf{k}}P({\bf k};{\bf q})G({\bf k})G^{*}({\bf k})P({\bf k};{\bf q})}\right)^{-1}
\eea
{where the ratio of two integrals is the analog of the $\langle \cos \theta \rangle$ term in the diffusion of electromagnetic waves \cite{KvTRMP}}. As a consequence, the diffusion constant can be represented as
\beq\label{diffin}
D =
\frac{B^{2}}{q^{2}\omega\left(1+a\right)}
\int\limits_{\bf{k}}\alpha P({\bf k};{\bf q})
G({\bf k})G^{*}({\bf k})P({\bf k};{\bf q})
\eeq
It must be noted here that $\alpha$ included in the general expression for the diffusion constant from  Eqn. (\ref{diffin}) can be evaluated explicitly using the
symmetry properties of the Green function, the self-energy, and the WTI from Eqns. (\ref{green}), (\ref{massop}),and (\ref{WLWTIzero}), respectively, as well as the reciprocity property of the kernel $\mathcal{K}$.
Indeed, those equations support the validity of the following relations:
\begin{eqnarray}\label{WTIzeroproof}
\Delta \Sigma(-\mathbf{k}^{\prime};{\bf 0},0)=\\ \int\limits_{\mathbf{k}^{\prime\prime}}\Delta G(\mathbf{k}^{\prime\prime};{\bf 0},0)\mathcal{K}^{B}(\mathbf{k}^{\prime\prime},-{\bf k}^{\prime};{\bf 0},0)=\nonumber\\
\Delta \Sigma(\mathbf{k}^{\prime};{\bf 0},0)=\nonumber\\ \int\limits_{\mathbf{k}^{\prime\prime}}\Delta G(\mathbf{k}^{\prime\prime};{\bf 0},0)\mathcal{K}^{B}(\mathbf{k}^{\prime\prime},{\bf k}^{\prime};{\bf 0},0)=\nonumber\\
\int\limits_{\mathbf{k}^{\prime\prime}}\mathcal{K}^{B}(\mathbf{k}^{\prime},{\bf k}^{\prime\prime};{\bf 0},0)\Delta G(\mathbf{k}^{\prime\prime};{\bf 0},0)=\nonumber\\
\int\limits_{\mathbf{k}^{\prime\prime}}\mathcal{K}^{B}(-\mathbf{k}^{\prime},{\bf k}^{\prime\prime};{\bf 0},0)\Delta G(\mathbf{k}^{\prime\prime};{\bf 0},0)=\Delta \Sigma(-\mathbf{k}^{\prime};{\bf 0},0)\nonumber \, .
\end{eqnarray}
This yields
\begin{equation}
 \mathcal{K}^{B}(\mathbf{k}^{\prime},{\bf k}^{\prime\prime};{\bf 0},0)=\mathcal{K}^{B}(-\mathbf{k}^{\prime},{\bf k}^{\prime\prime};{\bf 0},0)=\mathcal{K}^{B}(\mathbf{k}^{\prime},-{\bf k}^{\prime\prime};{\bf 0},0)
\label{symmetryprop}
\end{equation}
so that
\begin{equation}\label{odd}
\mathcal{K}^{B}({\bf k},-{\bf k}^{\prime\prime})G(-{\bf k}^{\prime\prime})G^{*}(-{\bf k}^{\prime\prime})P(-{\bf k}^{\prime\prime};{\bf q})=
-\mathcal{K}^{B}({\bf k},{\bf k}^{\prime\prime})G({\bf k}^{\prime\prime})G^{*}({\bf k}^{\prime\prime})P({\bf k}^{\prime\prime};{\bf q}) \, .
\end{equation}
In turn, as one can easily see from the Eq.(\ref{angularproof}), the odd character of the function from the Eq.(\ref{odd}) immediately determines the value $\alpha=1$. Therefore, the diffusion constant from the Eq.(\ref{diffin}) can be brought into the form
\beq\label{DCFin}
D =
\frac{B^{2}}{q^{2}\omega\left(1+a\right)}
\int\limits_{\bf{k}}P({\bf k};{\bf q})
G({\bf k})G^{*}({\bf k})P({\bf k};{\bf q})
\eeq
Finally, exploiting the representation for $G({\bf k})G^{*}({\bf k})$ through Eqn.(\ref{greenexpl}), along with Eqns. (\ref{parameters}),(\ref{GFDiff}) and (\ref{aparaprox}) we can write
\beq
\label{DCFinExpl0}
D =
\frac{c_{T}^{4}}{2\omega^{3}\left(1+a\right)}\frac{K^{2}(K^{2})^{*}}{Im[K^{2}]}\approx
\frac{c_{T}^{4}}{v^{4}\left(1+a\right)}\frac{vl}{2}\approx \frac{c_{T}^{2}}{v^{2}}\frac{vl}{2}
\eeq
where
\beq\label{VelLen}
v=\frac{\omega}{Re[K]}, \hspace{1em} l=\frac{1}{2Im[K]} \, ,
\eeq
and we have used the approximations $K^{2}(K^{2})^{*}\approx \omega^{4}/v^{4}$, and $Re[K^{2}]\approx \omega^{2}/v^{2}$. {It is easy to check that these approximations hold within terms linear in the density $n$. }

To sum up, we have the following relation for the diffusion coefficient of anti-plane waves traveling through a maze of screw dislocations:
\beq
\label{DCFinExpl}
D = \frac 12 v^* l \, ,
\eeq
the usual form of diffusion coefficients, in terms of a transport velocity $v^* = c_T^2 /v$, where $c_T$ is the ``bare'' wave velocity, $v$ is the velocity of coherent waves, and $l$ a transport mean free path which in this case is equal to the attenuation length of the coherent waves.

\section{Discussion and conclusions}
\label{disc}
We have computed, Eqn. (\ref{DCFinExpl}), the diffusion coefficient for anti-plane elastic waves moving incoherently through many, randomly placed, screw dislocations in two dimensions. Although the procedure is based on a standard Bethe-Salpeter approach, a number of features of the calculation deserve to be pointed out.

The first is that  {we are studying the diffusive behavior of anti-plane waves of frequency $\om$ in a two dimensional continuum. The limit $\om \rightarrow 0$ must be a physically realizable limit in this context. However, unless the Peierls Nabarro (PN) force is considered, the scattering cross section for an elastic wave by a single dislocation diverges at low frequencies and the problem would not be well defined. Also, the introduction of a PN force allows for the regularization of an otherwise divergent mass operator for the coherent waves, through a renormalization of the PN force constant. In other words, the PN force provides a frequency scale that is essential to the formulation of the wave diffusion problem.}

A second important feature is the Independent Scattering Approximation (ISA). In that case both the mass operator and the irreducible kernel of the BS equation are proportional to the density of scatterers $n$, {and all higher order terms in $n$ have been omitted.} The mass operator can then be computed to all orders in perturbation theory, in which the perturbation is carried out for weak dislocation-wave coupling. Given the nature of the interaction, Eqn. (\ref{potanti}), this is the case for long wavelengths. The summation is possible due to the point-like nature of the interaction. Also, the fact that the interaction involves a gradient of a delta function is responsible for the pre-WTI that is needed in order to obtain the WTI. {This, in turn, depends on the fact that the interaction between dislocation and elastic wave is given by the Peach-Koehler force.}  {When the ISA is used, however, and the low frequency and long wavelength limits that are needed to make sense of a diffusion behavior are applied, it becomes necessary to impose $B=0$; that is, there is no viscous damping associated with the string dynamics. Since damping is associated with retardation effects, this restriction can be associated with the fact that the interaction of the string with the elastic wave is evaluated at the equilibrium position of the string thus neglecting retardation effects. It is conceivable that relaxing this condition could lead to a compensation with $B\ne 0$ terms in the dynamics that has been considered through the frequency dependence of the potential (\ref{potanti})}

{
In addition, the approximation $ | Im[K^{2}] | \ll | k^{2}-Re[K^{2}] | $ has been used, where $K$ is the effective wave number of the coherent waves. As discussed in Appendix \ref{APK}, this places a restriction on the regions of $(\om,k)$ space, where the function $\Phi$ is defined, in which a diffusion behavior occurs when $B=0$. Along the diagonal $k \sim k_\beta = \om/c_T$, this is  automatically satisfied for frequencies $\om$ that are small compared to the natural frequency of the oscillating dislocations, as well as small compared to the viscous damping. It is not satisfied for frequencies around the resonant frequency, and it is again satisfied for high frequencies, both for small and high dampings.} 

{
In terms of the dislocation parameters, the low density approximation has different implications for low and high frequencies $\om$. For low frequencies, the distance between dislocations has to be large enough so that the time it takes the bare wave to go from one to the next is large compared with the period of oscillation around the PN potential well minimum. For high frequencies, the distance between dislocations must be large compared to bare wavelength.
}

\subsection{Concluding remarks}
A Bethe-Salpeter approach as been used to study the behavior of incoherent anti-plane waves inside a two-dimensional elastic continuum populated by a random distribution of screw dislocations. A diffusive limit has been identified and the corresponding diffusion constant has been calculated.  A natural next step would be to consider whether the diffusion coefficient can vanish, leading to localization of the waves. Another, certainly, would be to use the techniques developed in this paper to address the more involved, but also more realistic, case of elastic waves in a three-dimensional elastic continuum with many, randomly placed, vibrating dislocation segments.

\acknowledgements
This work was supported by Fondecyt Grants 1130382, 1160823,  and ANR-CONICYT grant 38, PROCOMEDIA. A useful discussion with M. Riquelme is gratefully acknowledged.

\appendix
\section{Summation of the perturbation expansion for the mass operator}
\label{Apa}
Following the ISA we evaluate the mass operator as $\Sigma= n\int <t>d\vec X^{n}_0 $, where the average is over the Burgers vector. We consider screw dislocations, with a Burgers vector of fixed magnitude but randomly oriented. Dislocation position has been assumed to be uniformly distributed with density (number per unit surface) $n$.
We introduce the definition of the $t$-matrix in momentum space through \cite{RG}
\beq
t(\vec k,\vec k')=\int d\vec x d\vec x' e^{-\imath \vec k \cdot \vec x}t(x,x')e^{\imath \vec k' \cdot \vec x'}
\eeq
and its Born expansion
\beq
t(\vec k,\vec k')=t^{(1)}(\vec k,\vec k')+t^{(2)}(\vec k,\vec k')+t^{(3)}(\vec k,\vec k') \ldots
\label{tseries}
\eeq
The first Born approximation is easily computed:
\bea
t^{(1)}(\vec k,\vec k')&=&\int d\vec x e^{-\imath \vec k \cdot \vec x}\mu V(x)e^{\imath \vec k' \cdot \vec x}
\nonumber \\
 & =& -\mu{\mathcal A}^{n}\vec k\cdot\vec k'e^{\imath \left(\vec k'-\vec k \right)\cdot \vec X^{n}_0} .
 \label{TBorn}
\eea
{ T}he second Born approximation is defined as
\bea
t^{(2)}_{ik}(\vec k,\vec k')&=&\mu^{2}\int d\vec x \, d\vec x' e^{-\imath \vec k \cdot \vec x}V(x)G^0 (x-x') V(x') e^{\imath \vec k' \cdot \vec x'} \\
& = & \frac{(-\mu{\mathcal A}^{n})^2}{(2\pi)^2} \int d\vec x \, d\vec x'  d \vec q \, e^{-\imath \vec k \cdot \vec x} \frac{\partial}{\partial x_m} \delta (\vec x - \vec X^{n}_0) (-\imath q_m) e^{-\imath \vec q \cdot \vec X^{n}_0} G^0(\vec q ) e^{\imath \vec q \cdot \vec x'}   \nonumber \\
& & \hspace{1em} \times \frac{\partial}{\partial x'_p} \delta (\vec x' - \vec X^{n}_0) (\imath k'_p) e^{\imath \vec k' \cdot \vec X^{n}_0} \nonumber \\
& = & \frac{(-\mu{\mathcal A}^{n})^2}{(2\pi)^2} \int d\vec q \, k'_p k_m q_m q_p G^0(\vec q )e^{\imath \left(\vec k'-\vec k \right)\cdot \vec X^{n}_0} \nonumber \\
 & = & \frac{(-\mu{\mathcal A}^{n})^2}{4\pi}I\vec k\cdot\vec k' e^{\imath \left(\vec k'-\vec k \right)\cdot \vec X^{n}_0} \, . \nonumber
 \label{2born}
\eea
where ($k_\beta \equiv \omega/c_T$)
\beq
 \label{eye}
I \equiv \int dq \, q^{3}G^0(\vec q)=\frac{1}{\mu}\int dq\frac{q^3}{(q^2-k_\beta^2)}
\eeq

In turn, the third order Born approximation to the $T$ matrix can be determined from the relation
\beq
t^{(3)}(\vec k,\vec k')    =  \mu^{3}\int d\vec x \, d\vec x' d\vec x'' \; e^{-\imath \vec k \cdot \vec x} V(\vec x)  G^0(\vec x-\vec x') V(\vec x') G^0(\vec x'-\vec x'')V(\vec x'') e^{\imath \vec k' \cdot \vec x''}
\label{third term}
\eeq
which yields, explicitly,
\beq
t^{(3)}(\vec k,\vec k')   = \frac{(-\mu{\mathcal A}^{n})^3}{(4\pi)^2}I^2\vec k\cdot\vec k' e^{\imath \left(\vec k'-\vec k \right)\cdot \vec X^{n}_0} \, .
\label{power}
\eeq
It can be easily seen that the series (\ref{tseries}) for $t(\vec k,\vec k')$ is geometric and can therefore be summed to get
\bea
t(\vec k,\vec k')=t^{1}(\vec k,\vec k') {\frac{1}{1+\frac{\mu{\mathcal A}^{n}I}{4\pi}} ,}
\label{eq:t}
\eea
{Although $I$, given by (\ref{eye}), diverges, it can be regularized with a scheme similar to the one} used in the case of a random ensemble of edge dislocations in 3D medium \cite{RG}. The precise definition of the integral $I$ from Eqn. (\ref{eye}) is
\bea
 \label{eyeRG}
I &\equiv& \frac{1}{\mu}\lim\limits_{\eta\rightarrow0}\int dq\frac{q^3}{(q^2-k_\beta^2-\imath \eta)} \nonumber \\
& = & \frac{1}{\mu}\left[\mathcal{P}\int dq\frac{q^3}{(q^2-k_\beta^2)}+\imath \pi\int dq \, q^{3}\delta(q^2-k_\beta^2)\right] \\
& = &
I^{{\mathcal P}}+I^{{\mathcal R}} \nonumber.
\eea
where $\mathcal{P}$ denotes the principal value. We have then
\bea
 \label{eyeRGcomp}
I^{{\mathcal P}}=\Re [I] &\equiv& \frac{1}{\mu}\mathcal{P}\int dq\frac{q^3}{(q^2-k_\beta^2)}\\
 I^{{\mathcal R}}=\imath \Im [I] &\equiv& \imath \frac{\pi}{\mu}\int dq \, q^{3}\delta(q^2-k_\beta^2)\nonumber.
\eea
where we have used
\bea
 \label{PV}
\lim\limits_{\eta\rightarrow 0}\frac{1}{x-\imath \eta}=\mathcal{P}\frac{1}{x}+\imath \pi\delta(x).
\eea
The second integral appearing in  Eqn. (\ref{eyeRGcomp}) can be evaluated at once. It is
\bea
 \label{eyeRGImcomp}
I^{{\mathcal R}}=\imath \frac{\pi}{\mu}\int\limits_{-\infty}^{\infty} dq \Theta(q)q^{3}\delta(q^2-k_\beta^2)=\imath \frac{\pi}{\mu}\frac{k_\beta^2}{2},
\eea
where $\Theta(q)$ is Heaviside function.
On the other hand, the first integral in Eqn. (\ref{eyeRGcomp}) is divergent and the introduction of a short wavelength (upper limit) cut-off $\Lambda$ is needed. As a consequence, we have the following expression for $\Lambda>k_\beta$:
\beq
 \label{eyeRGRecomp}
I^{{\mathcal P}}(k_\beta,\Lambda)\equiv \frac{1}{\mu}\mathcal{P}\int\limits_{0}^{\Lambda} dq\frac{q^3}{(q^2-k_\beta^2)}=\frac{1}{2\mu}\left[\Lambda^{2}+k_\beta^2\ln\left(\frac{\Lambda^{2}}{k_\beta^2}-1\right)\right]  {\approx \frac{\Lambda^2}{2\mu} \, .}
\eeq
Finally, the self-energy  $\Sigma$ reads
\bea
\Sigma & = & n\int <t>d\vec X^{n}_0=n{<\frac{-\mu{\mathcal A}^{n}}{1+\frac{\mu{\mathcal A}^{n}I^{{\mathcal R}}}{4\pi}}>}k^{2}  \nonumber \\
 & = & {\frac{-n\mu^2 b^2 k^2}{M} \frac{1}{\om^2 -\om_{0R}^2 + i \left( \om \frac BM + \frac{\mu b^2 \om^2}{8Mc_T^2} \right)}}	\, .
\label{massop1}
\eea
 {with $\om^2_{0R} = \om_0^2 - \mu b^2 \Lambda^2/8\pi M$}.

{The integral (\ref{eyeRGRecomp}) nominally diverges because in continuum mechanics there is no intrinsic length scale. Consequently, all wave vectors, even very high ones, have to be integrated over in (\ref{eyeRGRecomp}). But this cannot be completely correct, since we are dealing with an approximation to a material that is made of atoms and molecules, and it does have a natural length scale, the interatomic distance. We take this indeterminacy into account through a renormalization of the frequency $\om_0$ to $\om_{0R}$.
}

 {In order for the geometric series to converge, it is needed that $|\mu{\mathcal A}^{n}I| < 4\pi$. For low frequencies this means that $(\mu b^2/8\pi M)^{1/2} \Lambda < \om_0$. This is the only possibility, actually, for the vanishing viscosity case, $B=0$, that is needed in order to have a WTI in the ISA.}

\section{Optical theorem}
\label{Apb}
We need to show
\beq
\label{b1}
\Delta \Sigma(\mathbf{k}^{\prime})=\int\limits_{\mathbf{p}^{\prime\prime}}\Delta G(\mathbf{p}^{\prime\prime}){\cal K}^{B}({\bf p}^{\prime\prime},{\bf k}^{\prime};{\bf 0},0)
\eeq
with
\beq
{\cal K}^{B}({\bf k},{\bf k}^{\prime};{\bf 0},0)=n <t({\bf k},{\bf k}^{\prime})t^{*}({\bf k}^{\prime},{\bf k})> \, ,
\eeq
a limiting form of (\ref{isabol}).

Consider the right-hand-side of (\ref{b1}), {together with Eqns. (\ref{deltaGdelta})  and (\ref{eq:t})}:
\bea
\label{wtiint}
\int\limits_{{\bf k}}\Delta G({\bf k}){\cal K}^{B}({\bf k},{\bf k}^{\prime};{\bf 0},0) & = &
{\cal V}\int\limits_{{\bf k}}k^{2}\delta\left(k^{2}-Re[K^{2}]\right)\left({\bf k}\cdot{\bf k}^{\prime}\right)^{2} \\
 & = & \frac{({k}^{\prime})^{2}{\cal V}}{8\pi}(Re[K^{2}])^{2}\nonumber \\
  & \approx & \frac{({k}^{\prime})^{2}{\cal V}\om^4}{8\pi c_T^4}  \\
  & = & -\frac{nk'^2 \om^2}{8} \frac{{\cal A}^2}{|{\cal D}|^2}
  \label{b5}
\eea
with ${\cal V} = -n\pi \mu^2 \langle |{\cal T}|^2\rangle / \rho^2 \om^2$. { In the third line we have used $(Re[K^{2}])^{2} \approx \omega^{4} / v^{4} \approx \omega^{4} / c^{4}$ to leading order in the density $n$, as discussed in Appendix \ref{APK}.}  For convenience, we reproduce here Eqn. (\ref{poles})
\beq
\label{polesap}
K^2  =  \frac{\omega^2}{c_T^2 (1+n{\cal T}) }
\eeq
Let us write ${\cal T} \equiv {\cal A}/{\cal D}$, with ${\cal D} \equiv {1 + (\mu {\cal A} I^{{\mathcal R}})/ 4\pi}$.  For the left-hand-side of (\ref{b1}) we have, from (\ref{deltaGdelta}) and (\ref{polesap})
\bea\label{SEDiffap}
\Delta\Sigma({\bf k};{\bf 0},0) & = & -\frac{Im[K^{2}]\omega^{2} k^{2}}{K^{2}(K^{2})^{*}} \\
 & = & -n k^2 c_T^2 \frac{\cal A}{|{\cal D}|^2} Im [{\cal D}] \nonumber \\
 & = & -\frac{n\mu k^2 c_T^2}{4\pi} \frac{{\cal A}^2}{|{\cal D}|^2} {Im [I^{{\mathcal R}}]}
\eea
which is equal to (\ref{b5}) {provided $B=0$} since, by (\ref{eyeRGImcomp}), ${Im [I^{{\mathcal R}}]} = \pi \om^2 /2\mu c_T^2$. Consequently  Eqn. (\ref{b1}) is satisfied  {to leading order in $n$, for small $n$}, as needed. {Notice the importance of being able to sum the complete series for the $t$ matrix in order to establish the result of this Appendix.}

\section{Discussion of the approximations used in this work}
\label{APK}
{
This paper involves waves in interaction with scatterers. The waves are characterized by their frequency $\om$, and the scatterers by their density $n$, resonant frequency $\om_0$ and viscous damping $B/M$. The computations that have been carried out involve approximations that restrict the values these parameters are allowed to have. This Appendix discusses this situation.
}

{
A central assumption of this paper is that there is a random distribution of scatterers and that each is an independent random variable (the Independent Scattering Approximation, ISA). This leads to expressions (\ref{massISA}) and (\ref{isabol}) for the mass operator $\Sigma$ and kernel ${\cal K}$, respectively, that are linear in the density of dislocations $n$. Keeping correlations would lead to higher order terms in $n$. As a consequence, only terms linear in $n$ must be kept in all expressions throughout. This has consequences that are explored below.
}

{
In addition, the inequality
\beq
| Im[K^{2}] | \ll | k^{2}-Re[K^{2}] | \, ,
\label{eqone}
\eeq
has been used in Appendix \ref{Apb}, to verify that the WTI holds within the ISA, and in Appendix \ref{esti} to compute part of the diffusion constant,
 with
\begin{eqnarray}
\label{eq2}
K &=& \frac{\omega}{c\sqrt{\left[1-\frac{\Sigma}{\rho c^{2}k^{2}}\right]}}= \frac{\omega}{c\sqrt{1+n{\cal T}} }  = \frac{k_\beta}{\sqrt{1+n{\cal T}} } \, ,\\
\Sigma & = & -n\mu {\cal T} k^2 \, , \nonumber\\
{\cal T} & \equiv &  {\frac{\mu b^2}{M}\frac{1}{\om^2 -\om_{0R}^2 +i \frac{\mu b^2}{Mc_T^2} \om^2}   } \, ,\nonumber
\end{eqnarray}
where  $M= (\mu b^2/4\pi c_T^2) \ln (\delta/\epsilon) \approx \mu b^2/ c_T^2$, and we shall use the last approximation for estimates.  {These expressions coincide with (\ref{eq:tee}) when $B=0$, as required by the ISA to the WTI. }The average  {that is implicit} in (\ref{eq2}) can be ignored since it involves $b^2$ and the only remaining randomness (after having averaged over position to get proportionality to dislocation density $n$) is in the sign of $b$.
}

{
We have, from (\ref{eq2}) and for small densities $n$,
\beq
K^2 = k_\beta^2 \frac{1}{1+ n{\cal T}} \approx k_\beta^2 (1- n{\cal T}) \, .
\eeq
Rewriting the inequality (\ref{eqone}) in terms of the attenuation length $l =1/( 2 Im[K])$ and coherent velocity $v= \om/Re[K]$, one obtains that the frequency $\om$ and wavenumber $k$ are restricted by either
\begin{eqnarray}\label{FrequencyI}
\omega\ll \frac{v}{l}\left(\sqrt{k^{2}l^{2}+\frac{1}{2}}-\frac{1}{2}\right)
\end{eqnarray}
or
\begin{eqnarray}\label{FrequencyII}
\omega\gg \frac{v}{l}\left(\sqrt{k^{2}l^{2}+\frac{1}{2}}+\frac{1}{2}\right) \, .
\end{eqnarray}
We expect the region of wavenumbers large compared to the inverse of the attenuation length, $kl \gg 1$, to be of special interest. In this case these restrictions become $\om \ll kv$, or $\om \gg kv$.
}

{Along the diagonal $k \sim \om/c_T = k_\beta$ in $(\om,k)$ space it is possible to be a little more precise. In this case we have
\bea
k^2 - Re[K^2] &=& n\, k_\beta^2 Re[{\cal T}] \\
Im[K^2] & = & -n\,  k_\beta^2  Im[{\cal T}]
\eea
and the inequality (\ref{eqone}) translates into
\beq
|Im[{\cal T}] | \ll | Re[{\cal T}] | \, .
\label{ineq2}
\eeq
The consequences of this restriction depend on the dislocation parameters $M$, $B$, $\gamma$, as well as frequency $\omega$. Notice that the dependence on the Burgers vector $b$ has dropped out under the approximation in (\ref{prefanti}). We consider various regimes for the frequency $\omega$.
}

In the case of small frequencies {$\om  \ll  \om_{0R}$ the restriction (\ref{ineq2}) is automatically satisfied, and the small density requirement $|n{\cal T}| \ll 1$ means $nc^2 /\om_{0R}^2 \ll 1$. That is, the period of oscillation of the screw dislocation around its (renormalized) NP minimum is small compared to the time the wave takes to go from one dislocation to the nearest one.
}

{
Near resonance, $\om \sim \om_{0R}$, ${\cal T}$ is pure imaginary and (\ref{ineq2}) cannot be satisfied.
}

{
For high frequencies $\om  \gg {\om_{0R}}$,  we have
\beq
{\frac{|Im[{\cal T}] |}{ | Re[{\cal T}] | } ={\frac{\mu b^{2}}{8 M c^{2}}}\sim\frac{1}{8} }
\eeq
{and} inequality (\ref{ineq2}) is satisfied, to the extent that $1\ll 8$. The condition $|n{\cal T}| \ll 1$  here means $n \ll k^2$, or, the distance between dislocations must be large compared to wavelength.
}

\section{Perturbation scheme for the spectral problem}
\label{pert}
To build up a system of equations for the determination of the diffusive pole structure we have to substitute the series from Eqn. (\ref{Pseries}) into Eqn. (\ref{Hom}) and gather together all terms of the same order, both in $\Omega$ and $\mathbf{q}$. Moreover, we assume that at every order of the perturbation scheme both WTI from Eqn. (\ref{WLWTI}) and symmetry constraints from (\ref{symmetry}) are valid. Omitting the $\Omega$ and $\mathbf{q}$ arguments for brevity, this yields
\begin{eqnarray}
\label{ApPSser}
\int\limits_{\bf{k}^{\prime\prime}}(H({\bf k},{\bf k}^{\prime\prime})+H^{1\Omega}({\bf k},{\bf k}^{\prime\prime})
+H^{1\mathbf{q}}({\bf k},{\bf k}^{\prime\prime})+H^{2\mathbf{q}}({\bf k},{\bf k}^{\prime\prime})+\dots)\\
\times (f({\bf k}^{\prime\prime})+f^{1\Omega}({\bf k}^{\prime\prime})
+f^{1\mathbf{q}}({\bf k}^{\prime\prime})+f^{2\mathbf{q}}({\bf k}^{\prime\prime})+\dots) & =  & \nonumber\\
 \hspace{-7em}   (\lambda^{1\Omega}+\lambda^{1\mathbf{q}}+\lambda^{2\mathbf{q}} +\dots)(f({\bf k})+f^{1\Omega}({\bf k})
+f^{1\mathbf{q}}({\bf k})+f^{2\mathbf{q}}({\bf k}) +\dots) & & \nonumber \, .
\end{eqnarray}
At first order in $\Omega$ and zero order in $\mathbf{q}$ Eqn. (\ref{ApPSser}) easily leads to Eqns. (\ref{system0}) and (\ref{Omega}) in the text. In a similar manner, collecting the  first order in $\mathbf{q}$ terms from Eqn. (\ref{ApPSser}) we obtain the following equation for $\lambda^{1\mathbf{q}}$
\beq
\label{ApPSser1q}
\int\limits_{\bf{k}^{\prime\prime}}(H({\bf k},{\bf k}^{\prime\prime})
f^{1\mathbf{q}}({\bf k}^{\prime\prime})+H^{1\mathbf{q}}({\bf k},{\bf k}^{\prime\prime})
f({\bf k}^{\prime\prime}))=
\lambda^{1\mathbf{q}}f({\bf k}) \, .
\eeq
Integrating (\ref{ApPSser1q}) over ${\bf k}$ cancels the contribution from the first term on its left hand side because of the WTI. So that, using (\ref{zeroorder}) we have
\beq
\label{ApPSser1q2}
\int\limits_{\bf{k}}\int\limits_{\bf{k}^{\prime\prime}}H^{1\mathbf{q}}({\bf k},{\bf k}^{\prime\prime})
\Delta G({\bf k}^{\prime\prime})=
\lambda^{1\mathbf{q}}\int\limits_{\bf{k}}\Delta G({\bf k})
\eeq
The left hand side of (\ref{ApPSser1q2}) is equal to zero because of the WTI written to first order in $\mathbf{q}$, as well as the odd in $\mathbf{k}$ character
of the tensor $P$ from (\ref{Hoperator}). Therefore, we obtain $\lambda^{1\mathbf{q}}=0$.

To complete the set of equations for the reconstruction of $\lambda_{0}\left(\mathbf{q},\Omega\right)$ we need $\lambda^{2\mathbf{q}}$. To second order in $\mathbf{q}$ Eqn. (\ref{ApPSser}) gives
\beq\label{ApPSser2q}
\int\limits_{\bf{k}^{\prime\prime}}(H({\bf k},{\bf k}^{\prime\prime})f^{2\mathbf{q}}({\bf k}^{\prime\prime})
+H^{1\mathbf{q}}({\bf k},{\bf k}^{\prime\prime})f^{1\mathbf{q}}({\bf k}^{\prime\prime}) +\mathbf{H}^{2\mathbf{q}}({\bf k},{\bf k}^{\prime\prime})f({\bf k}^{\prime\prime}))
=
\lambda^{2\mathbf{q}}f({\bf k}) \, .
\eeq
Then, Eqn. (\ref{system2q}) of the text is obtained integrating (\ref{ApPSser2q}) over ${\bf k}$ and using the  explicit form of the WTI at corresponding orders.
\section{Solution  for $f^{1\mathbf{q}}({\bf k})$}
\label{solu}
To find the solution for $f^{1\mathbf{q}}({\bf k})$ we have to modify accordingly Eqn. (\ref{system1q}). Indeed, using (\ref{zeroorder}) we can rewrite its second term as
\begin{eqnarray}\label{ApSolTran}
\int\limits_{\bf{k}^{\prime\prime}}H^{1\mathbf{q}}({\bf k},{\bf k}^{\prime\prime})B\Delta G({\bf k}^{\prime\prime}) & = &
\int\limits_{\bf{k}^{\prime\prime}}B\left(P(\mathbf{q};{\bf k}^{\prime\prime})\delta_{\mathbf{k}^{\prime\prime},\mathbf{k}}\Delta G({\bf k})\right.  \\
& & \hspace{2em} - \left.H({\bf k},{\bf k}^{\prime\prime})\Delta G^{1\mathbf{q}}({\bf k}^{\prime\prime})\right) \nonumber \\
 & = & B\int\limits_{\bf{k}^{\prime\prime}}H({\bf k},{\bf k}^{\prime\prime})
 [ \int\limits_{{\bf k}_{2}}\Phi({\bf k}^{\prime\prime},{\bf k}_{2})P(\mathbf{q};{\bf k}_{2})- \Delta G^{1\mathbf{q}}({\bf k}^{\prime\prime}) ]\nonumber \, .
\end{eqnarray}
Where the first equality is a consequence of the symmetry property from Eqn. (\ref{symmetry})  and applying the WTI to $H^{1\mathbf{q}}({\bf k},{\bf k}^{\prime\prime})$. The second equality is a result of substituting $\delta_{\mathbf{k}^{\prime\prime},\mathbf{k}}\Delta G({\bf k})$ by its value given by (\ref{BSfinal}).
Hence,
\beq\label{ApSol1q}
f^{1\mathbf{q}}({\bf k}^{\prime\prime})=
-B\left(\int\limits_{{\bf k}_{2}}\Phi({\bf k}^{\prime\prime},{\bf k}_{2})P(\mathbf{q};{\bf k}_{2})-\Delta G^{1\mathbf{q}}({\bf k}^{\prime\prime})\right)
\eeq
and
\beq\label{ApSoldeltaG}
\Delta G^{1\mathbf{q}}({\bf k})  =  {\bf q}\cdot\frac{\partial\Delta G({\bf k};{\bf q}^{\prime},0)}{\partial {\bf q}^{\prime}}|_{{\bf q}^{\prime}=0}=-\frac{1}{2\imath\rho}{\bf q}\cdot\frac{\partial \left(Re[G^{-}(\mathbf{k})]\right)}{\partial {\bf k}}
\eeq
\section{Calculation of $D_{\Delta G^{1\mathbf{q}}}$}
\label{esti}
Here we show that, in the appropriate limit spelled out below, $D_{\Delta G^{1\mathbf{q}}} = 0$. From Eqn. (\ref{Dconstant}) we have
\begin{eqnarray}\label{Gq}
D_{\Delta G^{1\mathbf{q}}} & = & \frac{-B^{2}}{q^{2}\omega\left(1+a\right)}\int\limits_{\bf{k}}P({\bf k};{\bf q})\Delta G^{1\mathbf{q}}({\bf k}) \\
&=& \frac{-\mu B^{2}}{2\rho^{2} q^{2}\omega\left(1+a\right)}\int\limits_{\bf{k}}
\mathbf{q}\cdot\mathbf{k}\frac{\partial \left(Re[G^{-}(\mathbf{k})]\right)}{\partial {\bf k}}\cdot{\bf q}
\end{eqnarray}
where Eqn. (\ref{Gq}) has been obtained using Eqns. (\ref{parameters}) and (\ref{ApSoldeltaG}).

In Eqn. (\ref{Gq}) we have to deal with the following integral ($s,t = 1,2$)
\begin{eqnarray}\label{ApEstInt}
\mathbb{J}^{st} & = &
\int\limits_{\bf{k}}
\left(\frac{\partial \left(Re[G^{-}]\right)}{\partial k_{t}}\right)k_{s}.
\end{eqnarray}
Then, using (\ref{green}) we can write
\begin{eqnarray}\label{ApEstInt2}
\mathbb{J}^{st}& = & \int\limits_{\bf{k}}
\left(\frac{2 k_{t}k_{s}\left(2k^{2}F^{2}(\omega,k)Im[K^{2}]-Re[K^{2}]F(\omega,k)\right)}{\rho\omega^{2}Im[K^{2}]}
\right) \\
& = & \int\limits_{-\infty}^{\infty}\delta_{st}\Theta(k)k^{3}dk
\left(\frac{ 2k^{2}F^{2}(\omega,k)Im[K^{2}]-Re[K^{2}]F(\omega,k)}{2\pi\rho\omega^{2}Im[K^{2}]}
\right).\nonumber
\end{eqnarray}
Where
\beq
\label{Ffunc}
F(\omega,k)=\left(\frac{Im[K^{2}]}{\left(k^2-Re[K^{2}]\right)^{2}+Im[K^{2}]^{2}}\right)
\eeq
 The expression for $\mathbb{J}^{st}$ in (\ref{ApEstInt2}) contains an ill-defined term in the integrand, proportional to $F^{2}(\omega,k)$, that can be regularized as follows \cite{mahan}: Introduce a new variable $x \equiv \left(k^2-Re[K^{2}]\right)$, and consider the following auxiliary integrals:
\begin{eqnarray}\label{AuxInt}
\int\limits_{-\infty}^{\infty}\left(\frac{Im[K^{2}]}{x^{2}+Im[K^{2}]^{2}}\right)dx=\pi,\\
\int\limits_{-\infty}^{\infty}\left(\frac{Im[K^{2}]}{x^{2}+Im[K^{2}]^{2}}\right)^{2}dx=\frac{\pi}{2Im[K^{2}]}.\nonumber
\end{eqnarray}
They show that it is possible to make the replacements
\begin{eqnarray}\label{Rep}
\left(\frac{Im[K^{2}]}{x^{2}+Im[K^{2}]^{2}}\right) & \longrightarrow & \pi\delta(x),\\
\left(\frac{Im[K^{2}]}{x^{2}+Im[K^{2}]^{2}}\right)^{2} & \longrightarrow & \frac{\pi\delta(x)}{2Im[K^{2}]}.\nonumber
\end{eqnarray}
which yield the same result after integration over $x$. Moreover, both replacements give the proper asymptotic behaviour in the low-density limit (our case) when $Im[K^{2}]\rightarrow 0$  {or, more precisely, when $| Im[K^{2}] | \ll | k^{2}-Re[K^{2}] |$}. Therefore, on the basis of Eqns. (\ref{Ffunc})-(\ref{Rep}) we have
\begin{eqnarray}\label{FReg}
F(\omega,k) & = & \lim\limits_{Im[K^{2}]\rightarrow 0}\pi\delta\left(k^2-Re[K^{2}]\right),\\
F(\omega,k)^{2} & = & \lim\limits_{Im[K^{2}]\rightarrow 0}\frac{\pi\delta\left(k^2-Re[K^{2}]\right)}{2Im[K^{2}]}\nonumber
\end{eqnarray}
Technically, Eqn. (\ref{FReg}) indicates that in all final expressions the limit ${Im[K^{2}]\rightarrow 0}$, {whose implications have been discussed in Appendix \ref{APK},} must be kept in mind. From Eqns. (\ref{Gq}), (\ref{ApEstInt}), (\ref{ApEstInt2}) and (\ref{FReg}) it follows that
\begin{eqnarray}\label{Dqeval}
\mathbb{J}^{st}=0\Rightarrow D_{\Delta G^{1\mathbf{q}}}=0 \, .
\end{eqnarray}


\end{document}